\documentclass[aps,prd
,preprint,tightenlines,nofootinbib,showpacs]{revtex4}
\usepackage{amssymb,latexsym}
\usepackage{amsmath,amsbsy,bbm}
\usepackage{epsfig,bm}
\usepackage{graphicx,comment}
\DeclareMathOperator{\tr}{tr}
\DeclareMathOperator{\Erfc}{Erfc}

\begin{document}
\def\a{{\alpha}}
\def\b{{\beta}}
\def\d{{\delta}}
\def\D{{\Delta}}
\def\e{{\varepsilon}}
\def\g{{\gamma}}
\def\G{{\Gamma}}
\def\k{{\kappa}}
\def\l{{\lambda}}
\def\L{{\Lambda}}
\def\m{{\mu}}
\def\n{{\nu}}
\def\o{{\omega}}
\def\O{{\Omega}}
\def\S{{\Sigma}}
\def\s{{\sigma}}
\def\th{{\theta}}

\def\ol#1{{\overline{#1}}}

\def\Dslash{D\hskip-0.65em /}
\def\Dtslash{\tilde{D} \hskip-0.65em /}

\def\CPT{{$\chi$PT}}
\def\QCPT{{Q$\chi$PT}}
\def\PQCPT{{PQ$\chi$PT}}
\def\tr{\text{tr}}
\def\str{\text{str}}
\def\diag{\text{diag}}
\def\order{{\mathcal O}}

\def\meff{{m^2_{\text{eff}}}}

\def\Meff{{M_{\text{eff}}}}
\def\cF{{\mathcal F}}
\def\cS{{\mathcal S}}
\def\cC{{\mathcal C}}
\def\cE{{\mathcal E}}
\def\cB{{\mathcal B}}
\def\cT{{\mathcal T}}
\def\cQ{{\mathcal Q}}
\def\cL{{\mathcal L}}
\def\cO{{\mathcal O}}
\def\cA{{\mathcal A}}
\def\cR{{\mathcal R}}
\def\cV{{\mathcal V}}
\def\cH{{\mathcal H}}
\def\cW{{\mathcal W}}
\def\cM{{\mathcal M}}
\def\cD{{\mathcal D}}
\def\cN{{\mathcal N}}
\def\cP{{\mathcal P}}
\def\cK{{\mathcal K}}
\def\Qt{{\tilde{Q}}}
\def\Dt{{\tilde{D}}}
\def\St{{\tilde{\Sigma}}}
\def\cBt{{\tilde{\mathcal{B}}}}
\def\cDt{{\tilde{\mathcal{D}}}}
\def\cTt{{\tilde{\mathcal{T}}}}
\def\cMt{{\tilde{\mathcal{M}}}}
\def\At{{\tilde{A}}}
\def\cNt{{\tilde{\mathcal{N}}}}
\def\cOt{{\tilde{\mathcal{O}}}}
\def\cPt{{\tilde{\mathcal{P}}}}
\def\cI{{\mathcal{I}}}
\def\cJ{{\mathcal{J}}}

\def\eqref#1{{(\ref{#1})}}

\preprint{UMD-40762-426}

\title{Hadrons in Strong Electric and Magnetic Fields}

\author{Brian C.~Tiburzi}
\email[]{bctiburz@umd.edu}

\affiliation{
           Maryland Center for Fundamental Physics,
	  Department of Physics,
	  University of Maryland,
	  College Park, MD 20742-4111, USA
}

\begin{abstract}
We use chiral perturbation theory to investigate
hadronic properties in strong electric and magnetic fields. 
A strong-field power counting is employed, and
results for pions and nucleons 
are obtained using Schwinger's proper-time method. 
In the limit of weak fields, we accordingly recover
the well-known one-loop chiral perturbation theory
results for the electric and magnetic polarizabilities
of pions and nucleons. 
In strong constant fields,
we extend the Gell-Mann--Oakes--Renner relation. 
For the case of electric fields, 
we find that non-perturbative effects result in hadron decay.
For sufficiently strong magnetic fields, the chiral analysis confirms that  
the nucleon hierarchy becomes inverted
giving rise to proton beta-decay.  
Properties of asymptotic expansions are explored
by considering weak field limits.
In the regime where the perturbative expansion breaks down, 
the first-order term gives the best agreement
with the non-perturbative result. 
\end{abstract}


\date{\today}

\maketitle

%
\section{Introduction}

The study of quantum field theories in the presence of external fields is a rich subject. 
In the pioneering work of Euler and Heisenberg, 
the one-loop effective action for QED 
was exactly computed for the special case of a constant electromagnetic field. 
This result was later formalized by Schwinger~\cite{Schwinger:1951nm}
in an essentially modern way. 
Such non-perturbative solutions provide important insight
into vacuum structure,
renormalization, 
and the behavior of pertubative approximations, 
see~\cite{Dunne:2004nc} for a review of effective actions
and more recent developments.

Probing a complicated theory like QCD with
external electromagnetic fields provides a lever-arm with which 
to study the modification of vacuum and hadron structure. 
This in turn gives insight into the underlying QCD dynamics. 
In external electromagnetic fields,
model field theories, and low-energy models of QCD
have been studied~\cite{Kawati:1983aq,Klevansky:1989vi,Suganuma:1990nn,Schramm:1991ex,Gusynin:1994re,Gusynin:1994xp,Babansky:1997zh},
as well as the QCD vacuum~\cite{Kabat:2002er,Miransky:2002rp}.
A striking fact about the QCD vacuum is that it 
spontaneously breaks chiral symmetry, 
which is evidenced by the non-zero 
value of the quark condensate. 
Because electromagnetic fields couple
directly to quarks,
the behavior of the condensate,
and hence the ground state of QCD,
will be affected by external fields. 
There have been numerous 
studies of the effects
of strong fields on the condensate.
Focusing on the model-independent
investigations using chiral perturbation theory, 
the calculation of the condensate in a constant magnetic field
at zero quark mass has been carried out~\cite{Shushpanov:1997sf,Agasian:1999sx},
and additionally calculations have included the effect of finite temperature~\cite{Agasian:2001ym}.
The external field dependence of the chiral condensate
has also been recently investigated in the more realistic scenario of 
non-vanishing quark mass in constant electromagnetic fields~\cite{Cohen:2007bt,Werbos:2007ym}.
As the condensate lowers the vacuum energy, it can be expected
that the addition of a magnetic field (which contributes positively to the action density, $\propto \vec{B}^2$) 
will strengthen spontaneous chiral symmetry breaking by increasing
the condensate.
On the other hand, the addition of an electric field 
(which contributes negatively to the action density, $\propto -  \vec{E}^2$, and has associated instabilities)
weakens chiral symmetry breaking. 
These heuristic expectations are verified by the explicit model-independent
calculations using chiral perturbation theory.

In this work, we consider the modification of hadron structure in strong fields. 
We use chiral perturbation theory to investigate pion and nucleon properties
in constant electric and magnetic fields. 
Chiral dynamics affords us the systematic and model-independent 
tool to investigate the modification of hadron structure in external fields. 
In electric and magnetic fields, 
we determine the effective mass of charged and neutral pions, 
as well as of the proton and neutron. 
The results for charged and neutral pions are shown to satisfy 
a generalized form of the Gell-Mann--Oakes--Renner relation. 
While the derived results are at a fixed order in the chiral expansion, 
they are non-perturbative in the strength of the external field. 
For sufficiently weak fields, the derived mass shifts of the pions 
and nucleons correctly reproduce the well-known electric and magnetic
polarizabilities from one-loop chiral perturbation theory. 
Expanding about the weak field limit allows us to investigate the behavior of perturbative approximations, 
and directly confront the breakdown of perturbation theory.
We see explicitly the behavior of perturbative approximations beyond
the expansion's regime of validity.
In this situation, 
the leading term gives the best approximation, 
which can be remarkably good. 
We speculate about the presence of this phenomenon in a few other systems.

Our study of hadrons in strong electric and magnetic fields
is additionally motivated by lattice gauge theory simulations~\cite{DeGrand:2006aa}.
The structure of the QCD vacuum and spectrum can 
be studied in the presence of electromagnetic 
fields using external field techniques with lattice QCD. 
For small field strengths, the response of hadrons
is governed by their multipole moments and multipole polarizabilities. 
These are coefficients of the first few terms in a weak-field expansion 
of a hadron's two-point function in the external field. 
The introduction of constant gauge fields on a torus, however, 
requires quantization conditions on the strength of the 
field~\cite{'tHooft:1979uj,vanBaal:1982ag,Smit:1986fn,Rubinstein:1995hc,Detmold:2008xk,Aubin:2008hz}.
To achieve the smallest values of constant fields, 
the generic quantization condition has the form
\begin{equation}
e a^2 F_{\mu \nu} = \frac{2 \pi n }{N_\mu N_\nu}
,\end{equation}
where $a$ is the lattice spacing, 
$F_{\mu \nu}$ 
represents the non-vanishing components of the field strength, 
$N_\mu$ 
the number of the lattice sites in the direction specified by $\mu$, 
and 
$n$ 
is an integer. 
Satisfying the quantization condition on 
current lattices, 
$N_j = 24$, 
inevitably leads to non-perturbative effects
from pion loops because the ratio 
$e F_{\mu \nu} / m_\pi^2$ 
is not considerably smaller than unity.
Furthermore as the lattice pion masses are
brought down to the physical point 
(at fixed lattice sizes), 
this ratio will exceed unity. 
Extraction of moments and polarizabilities
will then be obscured 
by non-perturbative effects.
To address these effects, 
we determine the behavior of pion and nucleon 
two-point functions in external fields to all orders
in the ratio of field strength to pion mass squared. 
Furthermore the study of hadronic two-point functions 
in strong external fields extends the method proposed in~\cite{Detmold:2006vu} 
for extracting charged particle properties from lattice QCD.

An outline of our presentation and key results are as follows. 
In Section~\ref{s:Pi}, we study the effects 
of strong electric and magnetic fields on pions. 
To begin, we review the chiral 
Lagrangian in Section~\ref{s:PiL}, 
and derive the charged pion propagator in 
a strong-field power counting.
One-loop computations are pursued 
in Sections~\ref{s:PiB} and \ref{s:PiE},
where we derive the shift in the effective
mass due to external magnetic and 
electric fields, respectively.
Expanding the non-perturbative results
in powers of the external field
allows us to explore the behavior of 
asymptotic expansions. 
We extend and demonstrate  
low-energy theorems
for pions in external fields
in Section~\ref{s:Condensate}.
The nucleon is the subject of Section~\ref{s:Nuc}. 
We review the chiral dynamics of nucleons and 
deltas in Section~\ref{s:NucL}.
Effects of strong magnetic fields on the nucleon
are investigated in Section~\ref{s:NucB}, 
where spin-dependent and spin-independent
contributions are computed at one-loop order. 
We confirm that the proton $\b$-decays in 
strong magnetic fields. Surprisingly the leading
and next-to-leading contributions to this 
effect arise from terms that are completely local in the effective theory. 
The study of the nucleon in an external electric field
is carried out in Section~\ref{s:NucE}.
Technical details concerning the computation of loop 
diagrams with mesons and nucleons 
using the strong-field power counting  
are contained in Appendix~\ref{s:nukeBcalc} and \ref{s:nukeEcalc}, 
while analytic continuation is addressed in Appendix~\ref{s:analcont}.
In Section~\ref{s:summy}, we end with concluding remarks,
and speculations about the breakdown of perturbative expansions.

%
\section{Pion in Strong Fields}
\label{s:Pi}

\subsection{Chiral Lagrangian}
\label{s:PiL}

To investigate the properties of hadrons in strong 
external fields, we turn to chiral perturbation theory~\cite{Gasser:1983yg}.
Chiral dynamics is based solely on the symmetries
of low-energy QCD; and, as such, provides a model-independent
description of hadrons in terms of a few parameters determined 
from phenomenology.
When electromagnetically gauged, the chiral theory
can be used to derive electromagnetic properties of hadrons.


In the absence of external fields and at zero quark mass,
the Lagrangian of QCD has a 
$\text{U}(2)_\text{L} \otimes \text{U}(2)_\text{R}$
chiral symmetry%
\footnote{
Here we consider only two light quark flavors.
There is empirical evidence from lattice QCD
simulations that three-flavor chiral expansions
are ill-fated~\cite{Lin:2007ap,Allton:2008pn,WalkerLoud:2008bp,Aoki:2008sm}.
This suggests that the strange quark mass is
possibly too large to be considered a perturbation
about the $\text{SU}(3)$ chiral limit.
Strange hadrons then must be considered as heavy 
external flavor sources in the relevant $\text{SU}(2)$ 
multiplets~\cite{Roessl:1999iu,Beane:2003yx,Tiburzi:2008bk}. 
}
that is reduced to $\text{SU}(2)_\text{L} \otimes \text{SU}(2)_\text{R} \otimes \text{U}(1)_\text{B}$
by the axial anomaly. 
The $\text{U}(1)_\text{B}$ symmetry corresponds 
to global vector symmetry for each flavor, 
and leads to baryon number conservation. 
The vacuum state of QCD spontaneously breaks
the two-flavor chiral symmetry:
$\text{SU}(2)_\text{L} \otimes \text{SU}(2)_\text{R} \longrightarrow \text{SU}(2)_\text{V}$. 
The resulting Goldstone manifold can be parametrized
by $\Sigma$ which is an element of the coset 
$\text{SU}(2)_\text{L} \otimes \text{SU}(2)_\text{R} / \text{SU}(2)_\text{V}$, 
and the Goldstone bosons, which are the pions, are
non-linearly realized in $\Sigma$:
\begin{equation}
\Sigma = \exp ( 2 i \phi / f) ,
\qquad
\phi = 
\begin{pmatrix}
\frac{1}{\sqrt{2}} \pi^0 & \pi^+ \\
\pi^- & - \frac{1}{\sqrt{2}} \pi^0
\end{pmatrix}
.\end{equation}
The dimensionful parameter $f$ 
arises from QCD dynamics,
$f \sim \L_\text{QCD}$; 
and, when weak interactions are considered,  
$f$ can be identified with 
the chiral limit value of the pion decay constant. 
In our conventions, we have $f = 130 \, \texttt{MeV}$.

Under a chiral transformation:
$L \in \text{SU}(2)_\text{L}$, $R \in \text{SU}(2)_\text{R}$, 
we have the transformation $\Sigma \to L \Sigma R^\dagger$. 
We can build an effective theory of low-energy pions
by forming the most general chirally invariant 
Lagrangian. Additionally we must take into account
sources of explicit chiral symmetry breaking. 
For our considerations, these are classical electromagnetic fields 
and  the quark masses.
The electromagnetic fields explicitly break the chiral symmetry
down to the product of diagonal subgroups:
$\text{SU}(2)_\text{L} \otimes \text{SU}(2)_\text{R}
\longrightarrow
\text{U}(1)_\text{L} \otimes \text{U}(1)_\text{R}$.
Thus in the absence of quark masses, only the neutral pion
is a Goldstone boson. 
The quark masses explicitly break chiral symmetry down 
to the vector subgroup. 
With non-vanishing quark masses and non-zero electromagnetic fields,
the resulting low-energy theory has only a 
$\text{U}(1)_\text{V} \otimes \text{U}(1)_\text{B}$
symmetry. 
Taking into account the pattern of spontaneous and explicit
symmetry breaking, 
the chiral Lagrangian has the form
\begin{equation}
\cL 
=
\frac{f^2}{8}
<  D_\mu \Sigma^\dagger D_\mu \Sigma >
-
\frac{\l}{2}
< m_Q ( \Sigma^\dagger  + \Sigma ) >
\label{eq:LO}
,\end{equation}
where we have used a bracketed notation 
to denote flavor traces: $< A > = \tr (A)$.
The quark mass matrix is given by
$m_Q = \diag(m_u, m_d)$,
and we choose to work in the limit of 
strong isospin symmetry, 
$ m_u = m_d = m$. 
The action of the covariant derivative $D_\mu$
is specified by
\begin{equation}
D_\mu \Sigma 
= 
\partial_\mu \Sigma
+ 
i e A_\mu 
[ \cQ, \Sigma]
,\end{equation}
where $\cQ$ is the quark electric charge matrix, 
$\cQ = \diag( 2/3, -1/2)$, 
$e$ the unit of charge, with $e>0$,
and $A_\mu$ is the gauge potential for the electromagnetic field.
The parameter $\l$ arises from 
QCD dynamics, $\l \sim \L_\text{QCD}^3$. 
Physically $\l$ is the value of the 
chiral condensate in the chiral limit
(technically it is the negative of the condensate).

In writing the Lagrangian, we have only included the lowest-order 
terms. These terms scale as $\cO(\varepsilon^2)$, where
$\e$ is a small dimensionless number.  
We assume the power counting
\begin{equation} \label{eq:PC}
\frac{k^2}{\Lambda^2_\chi} \sim \frac{m^2_\pi}{\L^2_\chi} \sim \frac{e F_{\mu \nu}}{\L_\chi^2} \sim \varepsilon^2
.\end{equation}
Here $\Lambda_\chi$ is the chiral symmetry breaking scale, 
$\L_\chi \sim 2 \sqrt{2}  \pi f$, 
$k$ represents a typical pion momentum, 
$m_\pi$ is the pion mass, 
and 
$F_{\mu \nu}$
is the electromagnetic field strength arising from the gauge potential, 
$F_{\mu \nu} = \partial_\mu A_\nu - \partial_\nu A_\mu$. 
We work with constant electromagnetic fields throughout.
The leading-order pion mass can be determined by expanding
the Lagrangian to quadratic $\phi$ order in zero field, 
\begin{equation} \label{eq:GMOR}
f^2 m_\pi^2 
= 
4 \l  m
.\end{equation}
This is the Gell-Mann--Oakes--Renner relation. We will 
return to its validity in the presence of external fields 
below in Section~\ref{s:Condensate}. 
Notice the counting of the field strength allows
for relativistic effects because 
$e F_{\mu \nu} / m_\pi^2 \sim \varepsilon^{0}$.
Consequently the leading-order propagator
for charged pions is not that for a relativistic free particle.
The Born couplings to the charge are non-perturbative.


We will focus on two cases for the external field. 
The first is a magnetic field 
$\bm{B} = B \hat{\bm{z}}$, 
which for concreteness we take to be
specified by the gauge potential 
\begin{equation} \label{eq:AB}
A_\mu = ( - B x_2, 0, 0, 0)
.\end{equation} 
Of course we could always make a 
gauge transformation, and arrive at
the same magnetic field. 
When possible, we will express our answers in terms of 
gauge invariant quantities.
The second case is an electric field
$\bm{E} = E \hat{\bm{z}}$. 
To work with the electric field, we always
start with a Euclidean space gauge potential, 
for concreteness take
\begin{equation}
\label{eq:AE}
A_\mu = (0,0, - \cE x_4, 0)
,\end{equation}
and address the analytic continuation
$\cE \to i E$. 
In Euclidean space, the difference between
magnetic and electric fields arises because 
of our choice of time direction. 
We will see this when we calculate nucleon
properties using the coordinate-space LSZ reduction in 
Sections~\ref{s:NucB} and \ref{s:NucE}.
In Minkowski space, by contrast, there is 
considerable difference between magnetic and electric 
fields.

%
\begin{figure}
\epsfig{file=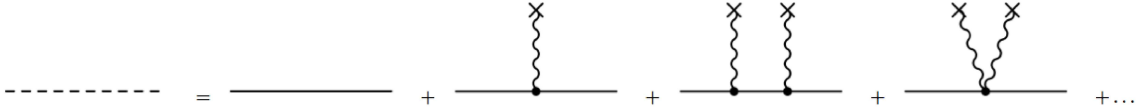,angle=270,width=15cm}
\caption{\label{f:prop} 
        Pion propagator in strong external fields. Depicted are the $\cO(\e^{2})$ Born couplings which must 
        be summed to derive the propagator in our power-counting scheme.}%
        \end{figure}
%

As mentioned above, the neutral pion propagator 
maintains a simple Klein-Gordon form. 
The charged pion propagator requires summation
of charge couplings, see Figure~\ref{f:prop}. 
This summation can be done by directly 
considering the charged pion action
Eq.~\eqref{eq:LO} at $\cO(\e^2)$:
\begin{equation}
\cL 
=
D_\mu \pi^- D_\mu \pi^+
+ 
m_\pi^2
\, 
\pi^- \pi^+
,\end{equation}
with 
$D_\mu = \partial_\mu + i e Q A_\mu$, 
where 
$Q = \pm 1$. 
Specializing to the case of the magnetic
gauge potential Eq.~\eqref{eq:AB}, 
we have
\begin{equation}
\cL 
= 
\pi^-
\left(
- \partial^2 
+ 
m_\pi^2
+
x_2^2 e^2 B^2
+
2 i x_2 e B \partial_1
\right)
\pi^+
.\end{equation}
Using the momentum mode expansion
\begin{equation}
\pi^+(x) 
= 
\int \frac{d \tilde{\bm{k}}} {(2 \pi)^3}
e^{ i \tilde{\bm{k}} \cdot \bm{x} } 
\pi^+_{\tilde{\bm{k}}} (x_2)
,\end{equation}
where $\tilde{\bm{k}}$ refers to 
the three non-vanishing components of
$\tilde{k}_\mu = (k_1, 0, k_3, k_4)$, 
the pion action becomes
\begin{equation}
S
= 
\int d x_2 
\int  \frac{d \tilde{\bm{k}}} {(2 \pi)^3}
\, 
\pi^-_{\tilde{\bm{k}}}(x_2) 
\left(
- \frac{\partial^2}{\partial x_2^2}
+
\tilde{\bm{k}}^2 + m_\pi^2
+
x_2^2 e^2 B^2
-
2 x_2 e B k_1
\right)
\pi^+_{\tilde{\bm{k}}}(x_2)
\label{eq:BSHO}
.\end{equation}
It will be easier to change variables:
\begin{equation}
X = x_2 - \frac{k_1}{e B}
.\end{equation}
Between states of good $\tilde{k}$, 
we need to find the propagator $D$, 
which as a concrete operator is
\begin{equation}
D^{-1} 
= 
2  \left( 
\frac{1}{2} 
p_X^2
+ 
\frac{1}{2}
e^2 B^2 X^2 
+
\frac{1}{2}
[\tilde{\bm{k}}^2 - k_1^2 + m_\pi^2]
\right)
\equiv
2   
( \cH  + E_\perp^2  /  2)
.\end{equation}
The $X$ and $p_X$ are canonically conjugate.
The transverse energy is defined by 
$E_\perp^2 
= 
k_3^2 + k_4^2 + m_\pi^2
$. 
The propagator 
$D(X',X)$ 
is then
\begin{equation}
D(X',X) 
= 
\langle X' | D | X \rangle
=
\frac{1}{2}
\int_0^\infty ds
\,
\langle X' | e^{ - s \cH} | X \rangle 
e^{ -  s E_\perp^2 /2 }
,\end{equation}
which can be thought of as a proper-time
evolution of a harmonic oscillator~\cite{Schwinger:1951nm}.
Given the form of the harmonic oscillator 
propagator
\begin{equation}
\langle X' , s | X , 0 \rangle 
= 
\sqrt{\frac{e B}{2 \pi \sinh e B s }}
\exp
\left\{
- \frac{eB}{2 \sinh e B s}
\left[
( X'^2 + X^2 ) \cosh e B s
- 2 X' X 
\right]
\right\}
,\end{equation}
we have thus derived the charged pion propagator
\begin{eqnarray} \label{eq:propB}
D(x',x)
&=& 
\frac{1}{2}
\int_0^\infty ds
\int \frac{d \tilde{\bm{k}}}{(2 \pi)^3}
e^{i \tilde{\bm{k}}  \cdot (\bm{x'} - \bm{x}) }
\Big\langle 
x'_2 - \frac{k_1}{ e B}, s 
\, \Big| \, 
x_2 - \frac{k_1}{ e B} , 0 
\Big\rangle
e^{ - s E_\perp^2 / 2} 
.\end{eqnarray}
We do not include an electromagnetic gauge link.%
\footnote{
Appending the Abelian gauge link, $L(x,0) = \int_0^x  d \zeta_\mu \exp [ i e A_\mu (\zeta) ]$,
leads to the modified Green's function relation
\begin{equation} \notag
\left( - \partial_\mu \partial_\mu + m_\pi^2 \right) L(x,0) D(x,0) = \delta^4(x)
.\end{equation} 
In this form, we see the action of the partial derivative is equivalent 
to the gauge covariant derivative. 
} 
The propagator in Eq.~\eqref{eq:propB} is a Green's function 
satisfying the relation
\begin{equation}
\left( - D_\mu D_\mu + m_\pi^2 \right) D(x,0) = \delta^4(x)
\label{eq:Green}
.\end{equation}
This relation will be useful in simplifying calculations.

The charged pion propagator in the case of a constant 
electric field, specified by Eq.~\eqref{eq:AE}, can be
written down from the magnetic result by inspection.
We find
\begin{eqnarray} \label{eq:propE}
D(x',x)
&=& 
\frac{1}{2}
\int_0^\infty ds
\int \frac{d \bm{k}}{(2 \pi)^3}
e^{i \bm{k} \cdot (\bm{x'} - \bm{x}) }
\Big\langle 
x'_4 - \frac{k_3}{ e \cE}, s 
\, \Big| \, 
x_4 - \frac{k_3}{ e \cE} , 0 
\Big\rangle
e^{ - s E_{k_\perp}^2 / 2} 
,\end{eqnarray}
with $E_{k_\perp}^2 = k_1^2 + k_2^2 + m_\pi^2$.
At this level, the only difference between propagators
is the particular labeling of coordinates.


Based on the power counting in Eq.~\eqref{eq:PC}, we need to include 
higher-order terms so that we can calculate amplitudes to one-loop level. 
Two further terms that depend on the external field strength 
must be added to the Lagrangian.%
\footnote{
In an electromagnetic field, where $\vec{E} \cdot \vec{B}$ is non-vanishing,
we would additionally need to consider terms from the Wess--Zumino--Witten
Lagrangian~\cite{Wess:1971yu,Witten:1983tw}. 
For a magnetic or electric field, such terms are absent unless
other sources, such as finite density, are considered, see~\cite{Newman:2005as,Son:2007ny}. 
} 
These higher-order terms scale generically as $\cO(\e^4)$, 
and will only contribute at tree-level. 
They have the form
\begin{equation}
\cL 
=
- i \a_9 F_{\mu \nu}
< \cQ D_\mu \Sigma D_\nu \Sigma^\dagger 
+ 
\cQ D_\mu \Sigma^\dagger D_\nu \Sigma >
- 
\a_{10}
F_{\mu \nu} F_{\mu \nu}
< 
\cQ \Sigma \cQ \Sigma^\dagger 
>
\label{eq:NLO}
.\end{equation}
These terms lead to  energy
shifts of charged particles 
proportional to the field strength squared.
Terms with higher powers of the 
external field are suppressed by the 
chiral symmetry breaking scale, and first occur 
in the $\cO(\e^6)$ Lagrangian. 
There is an additional field-dependent term 
that can potentially contribute from the fourth-order Lagrangian. 
This term is
\begin{equation}
\cL 
=
\a_4 \lambda
< D_\mu \Sigma D_\mu \Sigma^\dagger >\, <m_Q ( \Sigma + \Sigma^\dagger) >
.\end{equation}
Given the structure of this operator, 
it can be removed with a field redefinition of 
$\Sigma$ 
resulting in a multiplicative renormalization
of the mass term in Eq.~\eqref{eq:LO}. 
This renormalization does not depend on the 
external field strength, and thus only modifies
the bare pion mass. 
We can thus remove the $\a_4$ term in renormalized perturbation theory 
by working with the physical pion mass. 
Thus at $\cO(\e^4)$, the low-energy 
constants in Eq.~\eqref{eq:NLO} 
are the only new parameters required.

\subsection{Pion in Strong Magnetic Fields}
\label{s:PiB}

Having written down the relevant terms of the 
leading and next-to-leading order chiral Lagrangian, 
we can calculate pion two-point function at 
$\cO(\e^4)$. 
At this order, we have the tree-level couplings in 
Eq.~\eqref{eq:NLO} and one-loop diagrams. The
latter arise from expanding the leading-order 
Lagrangian Eq.~\eqref{eq:LO} to quartic order
in pion fields. 
%
\begin{figure}
\epsfig{file=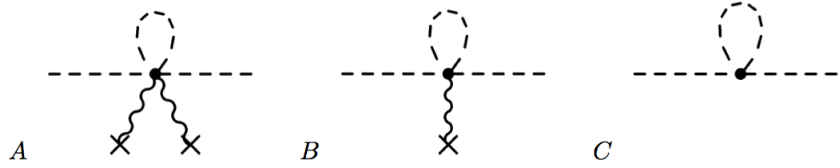,angle=270,width=11cm} 
        \caption{Diagrams for a pion in an external field.
        Shown are the one-loop diagrams for the pion 
        two-point function in an external field. The dashed
        lines represent the pion propagation in the background field. 
        The wiggly lines terminating in crosses represent 
        couplings of the vertices to the background field.}%
	\label{f:pion}
\end{figure}
%
There are three different diagrams generated at this order, 
and these are depicted in Figure~\ref{f:pion}. 
We will handle the simplest contribution first. 
Diagram $C$ receives contributions from the mass
term of the chiral Lagrangian Eq.~\eqref{eq:LO} expanded
to fourth order in the pion fields. 
The value of the tadpole diagram is proportional to 
the self-contracted propagator, which for the charged 
pions we denote by $\cD(B)$. The self-contracted propagator
for the neutral pion is then simply $\cD(0)$. 
In dimensional regularization, 
the self-contracted propagator
is
\begin{equation} \label{eq:selflove}
\cD(B)
\equiv
D(x,x) 
= \frac{( 4 \pi \mu^2)^\epsilon}{(4 \pi)^2}
\int_0^\infty ds \, s^{\epsilon -1} e^{ - s m_\pi^2} \frac{ e B}{\sinh e B s}
.\end{equation}
The result of the mass insertion vertex is a shift of the 
leading-order masses
\begin{eqnarray}
m_{\pi^\pm}^2 = m_\pi^2   &\longrightarrow& m_\pi^2  \left( 1 - \frac{1}{3f^2} [4 \cD(B) + \cD(0) ]  \right), \\
m_{\pi^0}^2 = m_\pi^2      &\longrightarrow& m_\pi^2  \left( 1 - \frac{1}{3f^2} [2 \cD(B)+ 3 \cD(0) ] \right) 
.\end{eqnarray}

The remainder of the loop contributions to the pion two-point function arise 
from the kinetic term in Eq.~\eqref{eq:LO}. Expanded to fourth order, the interaction
vertex consists of four pion fields, two with gauge covariant derivatives and two without.
Diagrams $A$--$C$ are related by gauge invariance; their contributions should be accordingly
grouped in a gauge covariant fashion.
A particularly useful simplification to note is that there are no loop contractions of
$D_\mu \phi$ with $\phi$.  This is obvious for the neutral pion because the 
covariant derivative is simply a partial derivative and the loop integration must 
vanish by Lorentz invariance:
\begin{equation}
\int d^d k \, \frac{k_\mu}{k^2 + m_\pi^2} = 0
.\end{equation}
For the charged pions, one must use the explicit
form of the propagator Eq.~\eqref{eq:propB} to see that such contractions also vanish. 
Basically the covariant derivative of the self-contracted propagator ends up without  
coordinate dependence, hence has no four-direction in which to point.
Thus there are only two ways the covariant derivatives can act to produce a nonzero
result: either both act on the fields contracted to the external legs,
or both act inside the loop.

Consider contractions where both covariant derivatives act on the external legs. 
In this case, the remaining two pion fields are contracted to form the tadpole
which by Eq.~\eqref{eq:selflove} is independent of the vertex location. 
The modification $\d G(x',0)$ to the two-point function thus has the form
\begin{equation}
\d G(x',0) = - C \, \cD(B) \int d^4 x \, D_\mu D(x',x) D_\mu D(x,0)
,\end{equation}
where $C$ is the overall numerical factor, and $D_\mu$ acts on the $x_\mu$-coordinates.
We can integrate by parts to relocate the covariant derivative from the final
state to the initial state. This is possible for the gauge field piece of the derivative,
because in the contraction either the charge is zero for the neutral pion; or, for the charged pion,
there must be a relative sign between the charges of the fields contracted to the external legs. 
Hence this correction to the two-point function has the form
\begin{eqnarray} \label{eq:correction}
G(x',0) 
&=& 
D(x',0) 
- 
C  \, \cD(B) 
\int d^4 x \,  D(x',x)  
\left( - D_\mu D_\mu \right) D(x,0)
.\end{eqnarray}
At this point, we could use the Green's function relation in Eq.~\eqref{eq:Green}, 
but it is best for later purposes to expose that we are calculating the wavefunction 
renormalization factor. 
To this end, we write Eq.~\eqref{eq:correction} in an operator notation
\begin{eqnarray}
G(x',0) 
&=& 
\langle x' | D | 0 \rangle
- C\, \cD(B)
\int d^4 x \, \langle x' | D | x \rangle \left( - D_\mu D_\mu \right)  \langle x | D | 0 \rangle 
\notag \\
&=&
\langle x' |  D - C \, \cD (B) D ( - D_\mu D_\mu )  D | 0 \rangle 
= 
\langle x' |
 D \frac{1}{1 + C\, \cD(B) (- D_\mu D_\mu ) D} | 0 \rangle
,\end{eqnarray}
where, in the last line, we treat $\cD(B)$ as a perturbation. 
Using the identity
\begin{equation}
\frac{1}{1 + A D}
=
D^{-1} 
\left[ 
D^{-1} + A
\right]^{-1}
,\end{equation}
we find
\begin{equation}
G(x',0) 
=
\langle x' | \frac{1}{D^{-1} + C \, \cD(B) ( - D_\mu D_\mu ) } | 0 \rangle
.\end{equation}
As an operator, $D^{-1} = - D_\mu D_\mu + m_\pi^2$, leading us to 
the final form for the correction
\begin{equation}
G(x',0) 
=
\langle x' | \frac{1}{[1 + C \, \cD(B)] ( - D_\mu D_\mu )  + m_\pi^2} | 0 \rangle
.\end{equation}
The factor $[1 + C \, \cD(B)]$ is just the wavefunction renormalization $Z$. 
Rescaling the pion fields appropriately, we have
\begin{eqnarray}
\label{eq:piplusZ}
\pi^{\pm} 
&\longrightarrow& 
\pi^\pm Z_{\pi^\pm}^{-1/2} 
=
\pi^\pm \left( 1 + \frac{1}{3f^2}  [ \cD(B) + \cD(0) ] \right),
\\
\label{eq:pizeroZ}
\pi^0 
&\longrightarrow&
\pi^0 Z_{\pi^0}^{-1/2} 
=
\pi^0 \left( 1 + \frac{2}{3 f^2} \cD(B) \right)
.\end{eqnarray}

The final loop contribution arises from contractions where both
covariant derivatives act inside the loop. In the case of the 
neutral pion, we have the dimensionally regulated loop integral
\begin{equation} \label{eq:neutralpion}
\int d^d k \,  \frac{k^2}{k^2 + m_\pi^2} = - m_\pi^2 \cD(0)
.\end{equation}
Essentially the same relation holds for the charged pions
which we can demonstrate using an integration by parts. 
Using brackets between fields to denote contractions, 
this contribution from the vertex has the form
\begin{eqnarray} 
\int d^4 x \, \phi [ D_\mu \phi D_\mu \phi ] \phi
= 
\int d^4 x
\Big(
\phi [ - D_\mu D_\mu \phi \, \phi ] \phi 
-
\partial_\mu [ \phi D_\mu \phi] \phi
- 
\phi [ \phi D_\mu \phi ] \partial_\mu \phi
\Big)
.\end{eqnarray}
Loop contractions of $D_\mu \phi$ with $\phi$, however, vanish
and the second two terms are hence zero. 
On the first term we use the Green's function relation Eq.~\eqref{eq:Green} 
and hence charged pion loop contributions from this term 
are the natural generalization of Eq.~\eqref{eq:neutralpion}: $ - m_\pi^2 \cD(B)$. 
The flavor and overall normalization factors are the same as the wavefunction
renormalization, hence that contribution is effectively doubled.

To present the final results, we must renormalize our expressions. 
We choose to renormalize in terms of the physical pion mass. 
The net effect of this renormalization is a subtraction of the zero-field
results. The self-contracted propagator becomes
$\ol \cD(B)  \equiv \cD(B) - \cD(0)$, which is free of divergences.  
This function can be calculated in closed form
\begin{equation}
\ol \cD(B) = \frac{e |B|}{(4 \pi)^2} \mathcal{I} \left( \frac{m_\pi^2}{ e |B|} \right)
,\end{equation}
where~\cite{Cohen:2007bt}
\begin{eqnarray}
\mathcal{I}(x) 
&=&
\int_0^\infty \frac{ds}{s^2}
e^{- x s}
\left( \frac{s}{\sinh s} - 1 \right)
=
x \left( 1 - \log \frac{x}{2} \right) + 2 \log \Gamma \left( \frac{1 + x}{2} \right) - \log 2 \pi
.\end{eqnarray}

Assembling the tree-level and loop contributions, 
the charged pion propagator 
at one-loop order has the form of $D(x',x)$ in Eq.~\eqref{eq:propB},
up to the replacement
\begin{equation}
m_\pi^2 
\longrightarrow 
m^2_{ \text{eff}, \pi^\pm }
= 
m_\pi^2 
+
\frac{1}{f^2} 8 e^2 B^2  ( \alpha_9 + \alpha_{10} )
.\end{equation}
Notice that the loop contributions have exactly 
cancelled at this order. 
In net, the charged pion two-point function
only receives corrections from neutral pion loops
and these are unaffected by the background field.%
\footnote{
This accident is additionally true to one-loop order 
in $SU(3)$ chiral perturbation theory
as well. 
In that case, the charged pion two-point function  
only receives loop contributions
from the neutral pion, and the eta.
} 
The term quadratic in the external field
is proportional to the magnetic polarizability, 
i.e.
\begin{equation}
m_{\text{eff}} (B) = m - \frac{1}{2} 4 \pi \beta_M  B^2 + \cO(B^4) 
.\end{equation}
Comparing with this expression, we find the known 
next-to-leading order result~\cite{Holstein:1990qy}
\begin{equation}
\b_M^{\pi^\pm} 
= 
- 8 \, \a_{f.s.}   \frac{\a_9 + \a_{10}} {f^2 m_\pi}
,\end{equation}
with the fine structure constant $\a_{f.s.} = e^2 / 4 \pi$.%
\footnote{
Results for the electric and magnetic polarizabilities of both charged
and neutral pions are known at next-to-next-to-leading order in the 
chiral expansion~\cite{Bellucci:1994eb,Burgi:1996qi,Gasser:2005ud,Gasser:2006qa}. 
Only at two-loop order would we recover these results in 
a weak field expansion. 
}
Terms with $B^4$ or higher powers of $B$ in the effective
mass squared arise at two-loop order, and
are hence suppressed by $m_\pi^2 / \L_\chi^2$.

For the neutral pion, however, the situation is quite the opposite. 
There is no local contribution at next-to-leading order,
and the loop diagrams do not exactly cancel. The neutral
pion propagator at one loop maintains a Klein-Gordon form,
but with an effective mass squared given by
\begin{equation} \label{eq:nonpert}
m^2_{ \text{eff}, \pi^0 }
= 
m_\pi^2 
\left[
1
+
\frac{2 e |B|}{( 4 \pi f)^2} 
\mathcal{I} \left( \frac{m_\pi^2}{e |B|} \right)
\right]
.\end{equation}
Taking the non-relativistic limit, 
$e |B| \ll m_\pi^2$, we have the perturbative
expansion of 
neutral pion energy
in a weak magnetic field
\begin{equation}
E_{\pi^0}(\bm{p} = \bm{0})
=
m_\pi
+ 
\frac{m_\pi^3}{(4 \pi f)^2}
\left[
- \frac{1}{6} \left( \frac{e B}{m_\pi^2} \right)^2
+ \frac{7}{180}  \left( \frac{e B}{m_\pi^2} \right)^4
- \frac{31}{630}  \left( \frac{e B}{m_\pi^2} \right)^6
+ \ldots 
\right] 
.\end{equation}
To derive this expansion, we utilized the integral
representation of the function $\mathcal{I}(x)$,
and series expanded the integrand. 
This yields only an asymptotic series
because the integral does not converge uniformly. 
The first term in this asymptotic series is the energy
shift of the neutral pion due to the magnetic polarizability, 
which we can read off as
\begin{equation}
\beta_M^{\pi^0} 
= 
\frac{\a_{f.s}}{ 3 ( 4 \pi f)^2 m_\pi}
.\end{equation}
This agrees with the known one-loop result~\cite{Bijnens:1987dc,Donoghue:1988ee,Holstein:1990qy}.
Unlike the charged pion, there are terms of all orders in $ e |B| / m_\pi^2$
generated in the neutral pion energy at one loop. 
These terms correspond to irreducible $\pi^0$-multiphoton interactions.

We can explore the behavior of the asymptotic expansion compared
to the non-perturbative result.
To do this,  we consider the relative change
in the neutral pion energy $\delta E$ defined by
\begin{equation} \label{eq:deltaE}
\delta E \equiv \frac{( 4 \pi f)^2}{m_\pi^2 }
\left[
\frac{E_{\pi^0}(\bm{p} =\bm{0}) - m_\pi}{m_\pi}
\right]
.\end{equation}
The exact numerical factor for the expansion parameter is arbitrary. 
We choose
\begin{equation} \label{eq:xi}
\xi = \frac{e|B|}{\sqrt{6} m_\pi^2}
,\end{equation}
so that the first two terms of the asymptotic expansion are of order unity, 
\emph{viz}.
\begin{equation}  \label{eq:pert}
\delta E = - \xi^2 + \frac{7}{5} \xi^4 - \frac{372}{35} \xi^6 + \ldots
.\end{equation}
Because the series is asymptotic, large coefficients will automatically appear
at some order in the expansion. 
%
\begin{figure}
\epsfig{file=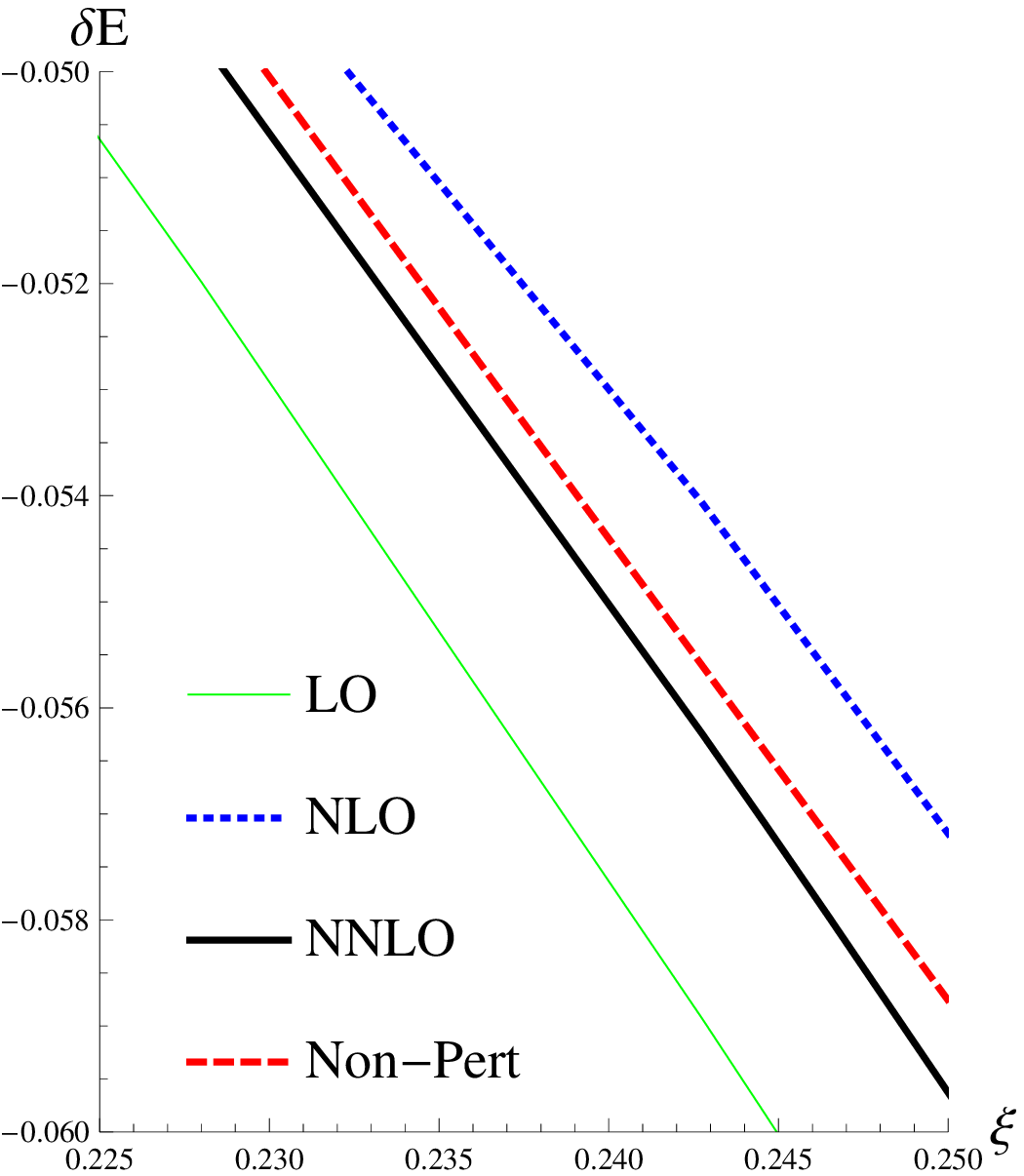,width=5.cm} 
\epsfig{file=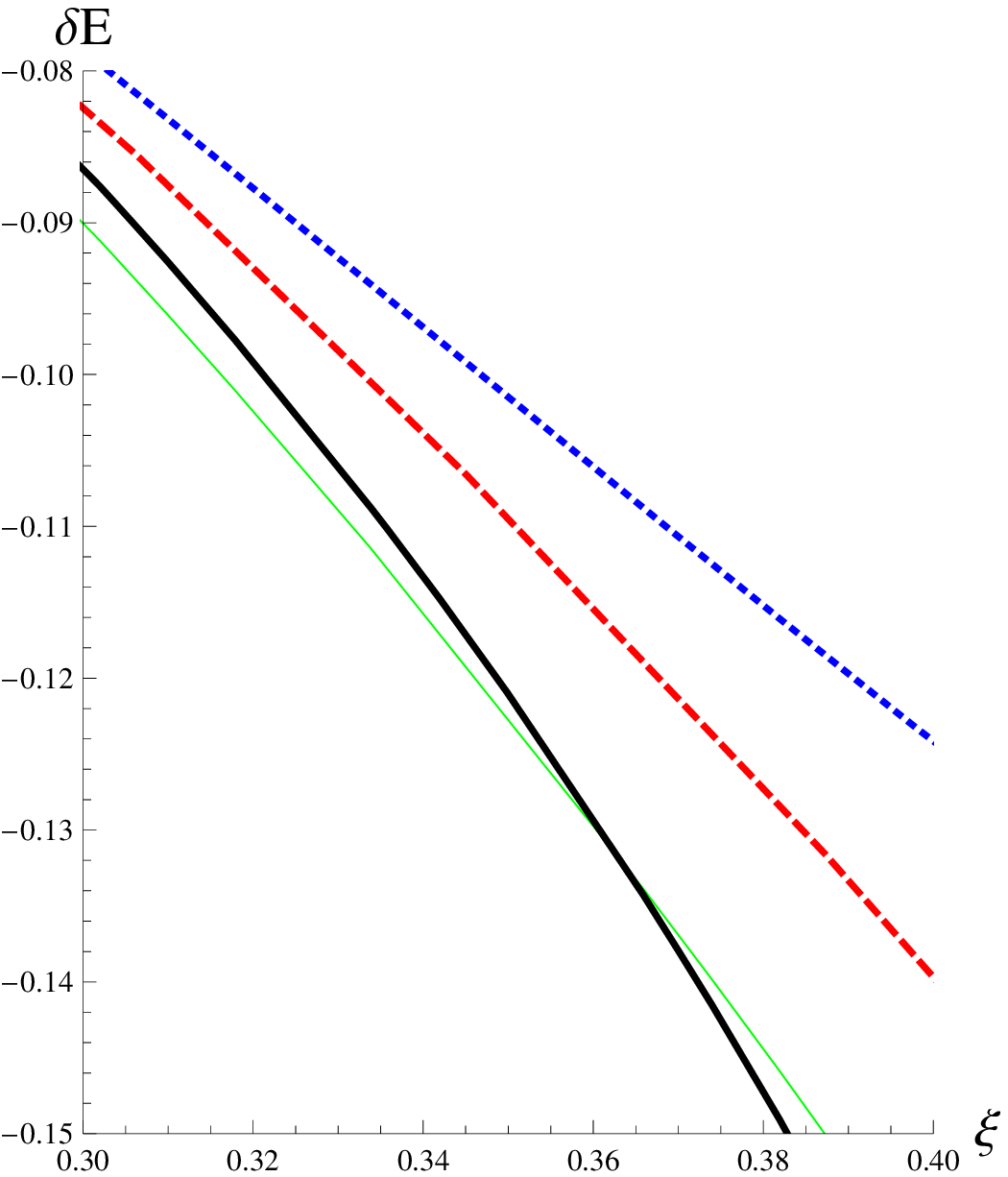,width=5.cm} 
\epsfig{file=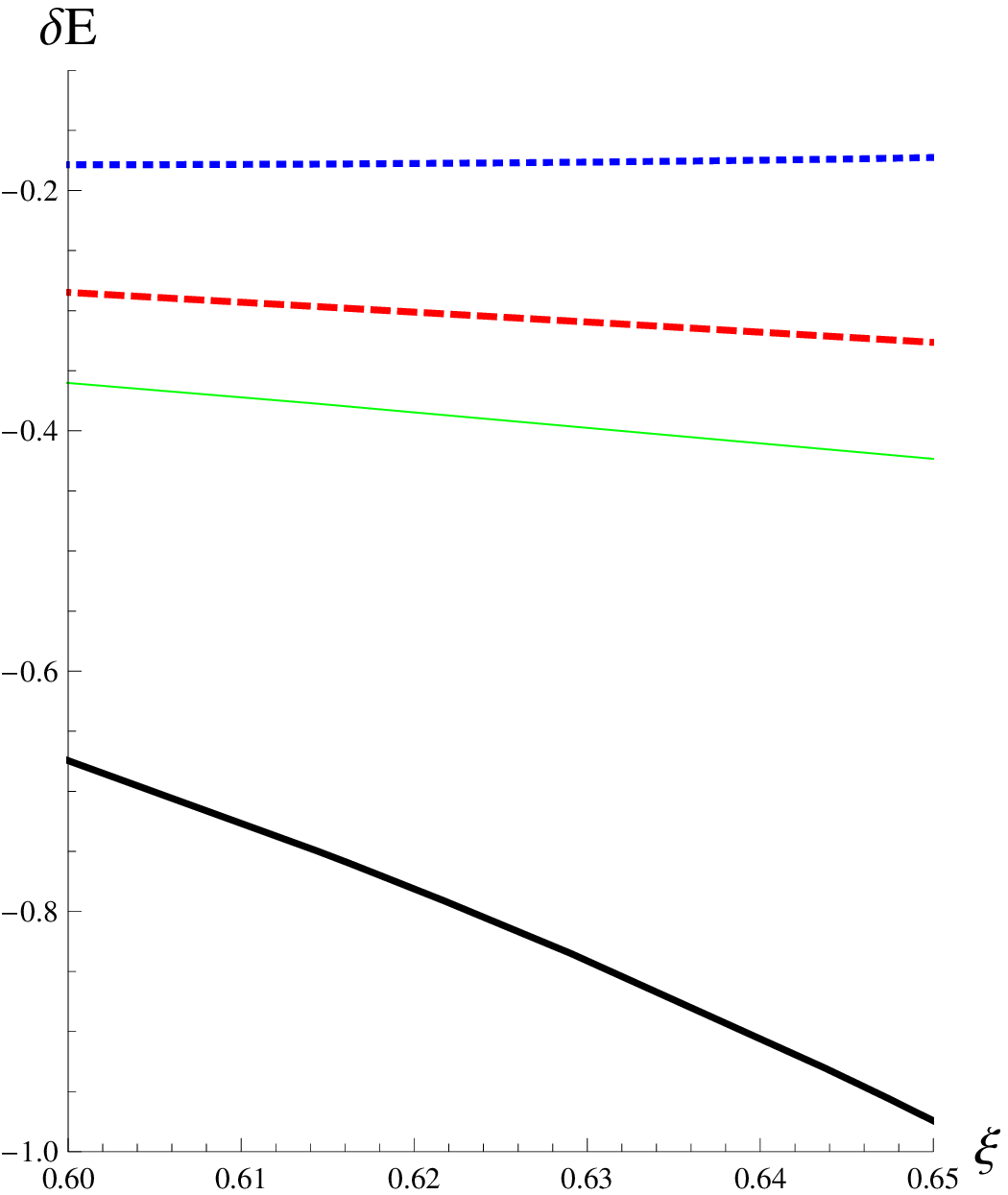,width=5.cm} 
        \caption{Comparison of perturbative and non-perturbative results in constant magnetic fields.
        Plotted is the change in pion energy $\d E$ in Eq.~\eqref{eq:deltaE}
        as a function of the expansion parameter $\xi$, Eq.~\eqref{eq:xi}. 
        LO, NLO, and NNLO are used to denote the order of the perturbative approximation in field strength.
        From left to right, 
        the three panels depict the perturbative regime, the transition regime, 
        and the non-perturbative regime. }%
	\label{f:compare}
	\end{figure}
%
We can investigate the nature of the asymptotic series by comparing
the non-perturbative result for $\d E$, in Eq.~\eqref{eq:deltaE}, 
to the perturbative series, Eq.~\eqref{eq:pert}. 
This is done in Figure~\ref{f:compare}, where we compare the leading-order
(LO), next-to-leading-order (NLO), and next-to-next-to-leading-order (NNLO)
results for the energy shift with the non-perturbative result. 
Notice we are working to a fixed order in the chiral expansion, 
and the expansion we are considering is in the strength of the external field.
Three different regimes of the expansion are shown. 
In the perturbative regime, the expansion parameter $\xi \sim 1/4$,
and the higher-order results better approximate the non-perturbative answer.
In the transition regime, the expansion parameter $\xi \sim 1/3$,
and adding higher-order terms is no longer efficacious. 
There is a point where the LO, NLO, and NNLO results
all differ by the same amount (in magnitude) from the non-perturbative result.
This point marks the transition from the perturbative regime to the non-perturbative
regime. 
The non-perturbative regime is typified for $\xi \sim 2/3$. 
In this region, all higher-order terms push the result 
away from the non-perturbative answer. 
Consequently the LO result agrees best,
$\sim 20 \%$, but cannot be improved using perturbation theory. 
This behavior is precisely as one expects  from an asymptotic  series:
only the first few terms give a good approximation. Beyond 
these terms, additional corrections are uncontrolled. 
When the expansion parameter is large, we are necessarily 
restricted to just the LO term.

%
\begin{figure}
\epsfig{file=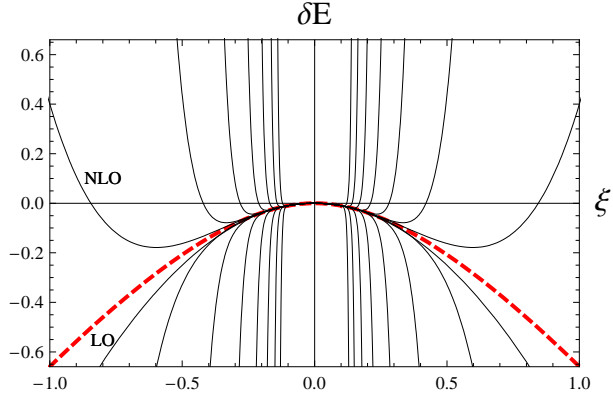,width=8cm} 
        \caption{Further comparison of perturbative and non-perturbative results in constant
        magnetic fields.
        Plotted is the change in the pion energy $\delta E$ vs.~$\xi$. The dashed curve
        shows the non-perturbative result from Eq.~\eqref{eq:nonpert}.
        The solid curves show the perturbative results to a given order, up to fifteenth
        order in the field-strength squared. The higher-order results accumulate near $\xi=0$.
                }
	\label{f:PvsNP}
\end{figure}
%
Finally in Figure~\ref{f:PvsNP}, we compare perturbative
results up to fifteenth order in the expansion. 
For a given order in the expansion, there is a value
for the expansion parameter $\xi$ for which the 
perturbative result diverges from the non-perturbative curve. 
The oscillatory nature of the perturbative expansion can
be seen from the plot, as all odd-order approximations undershoot
the non-perturbative result, while the even-order approximations overshoot 
the non-perturbative result. 
The region of $\xi$ where n${}^{\text{th}}$-order result gives
a reasonable description of the exact result 
is bounded in magnitude roughly by $1/n$. 
Consequently the higher-order curves accumulate on the plot
near $\xi = 0$, and 
reflect that the series expansion has strictly
zero radius of convergence.

\subsection{Pion in Strong Electric Fields}
\label{s:PiE}

It is straightforward to repeat the calculation of the
pion two-point functions at one-loop order in an external
electric field. The diagrams which contribute are identical
to those for the magnetic case. 
In general, we cannot write down the particle's energy shift
in an electric field from that in a magnetic field 
because the energy shift is not Lorentz invariant. 
The tadpole topology offers a valuable simplification, however, as
the tadpole loop-factor does not depend on the location of the pion 
source and sink. For example, we need not worry 
about the relative time entering the loop integration. 
Because we are afforded a Lorentz covariant treatment to 
arrive at the effective mass squared, the electric and magnetic results are consequently the same. 
This only occurs to one-loop order. 
(Moreover this accident does not occur for nucleons at one loop, 
as we shall see below in Section~\ref{s:NucE}.)

Given the simple tadpole loop topology at next-to-leading order, 
we can thus simply write down the Euclidean electric field results. 
For the charged pion, we have
\begin{eqnarray}
m_\pi^2 
\longrightarrow 
m^2_{ \text{eff}, \pi^\pm }
&=& 
m_\pi^2 
+
\frac{1}{f^2} 8 e^2 \cE^2  ( \alpha_9 + \alpha_{10} )
\notag \\
&=& 
m_\pi^2 
-
\frac{1}{f^2} 8 e^2 E^2  ( \alpha_9 + \alpha_{10} )
,\end{eqnarray}
where, in the second line, we have performed the trivial
analytic continuation to Minkowski space.
The term quadratic in the external field
is proportional to the electric polarizability, 
i.e.
\begin{equation}
m_{\text{eff}} (E) = m - \frac{1}{2} 4 \pi \alpha_E  E^2 + \cO(E^4) 
.\end{equation}
Comparing with this expression, we find the known 
next-to-leading order result~\cite{Holstein:1990qy}
\begin{equation}
\a_E^{\pi^\pm} 
= 
8 \, \a_{f.s.}   \frac{\a_9 + \a_{10}} {f^2 m_\pi}
,\end{equation}
and the relation: $\alpha_E + \beta_M = 0$, 
which is a consequence of the helicity structure of the 
one-loop graphs.

For the neutral pion, we can also write down the Euclidean
space results by inspection, namely
\begin{equation} \label{eq:nonpertE}
m^2_{ \text{eff}, \pi^0 }
= 
m_\pi^2 
\left[
1
+
\frac{2 e \cE}{( 4 \pi f)^2} 
\mathcal{I} \left( \frac{m_\pi^2}{e \cE} \right)
\right]
,\end{equation}
but the analytic continuation to Minkowski space
requires care. 
First let us note that the asymptotic expansion of 
$i \,  \cI (-i x)$ needed for the electric field case 
is just the trivial analytic continuation of the
Euclidean space asymptotic expansion. 
Subtleties in the analytic continuation can only 
show up in non-perturbative physics.
Thus we have the same perturbative 
expansion for the neutral pion energy as 
above under the simple replacement $B \to i E$
(or $\xi \to i \xi$). 
The neutral pion energy consequently
has the weak electric field expansion
\begin{equation}
E_{\pi^0}(\bm{p} = \bm{0})
=
m_\pi
+ 
\frac{m_\pi^3}{(4 \pi f)^2}
\left[
\frac{1}{6} \left( \frac{e E}{m_\pi^2} \right)^2
+ \frac{7}{180}  \left( \frac{e E}{m_\pi^2} \right)^4
+\frac{31}{630}  \left( \frac{e E}{m_\pi^2} \right)^6
+ \ldots 
\right] 
.\end{equation}
From this expression, we can read off the electric polarizability
\begin{equation}
\a_E^{\pi^0} 
= 
-
\frac{\a_{f.s}}{ 3 ( 4 \pi f)^2 m_\pi}
,\end{equation}
which is in agreement with the known one-loop result~\cite{Bijnens:1987dc,Donoghue:1988ee,Holstein:1990qy}.

%
\begin{figure}
\epsfig{file=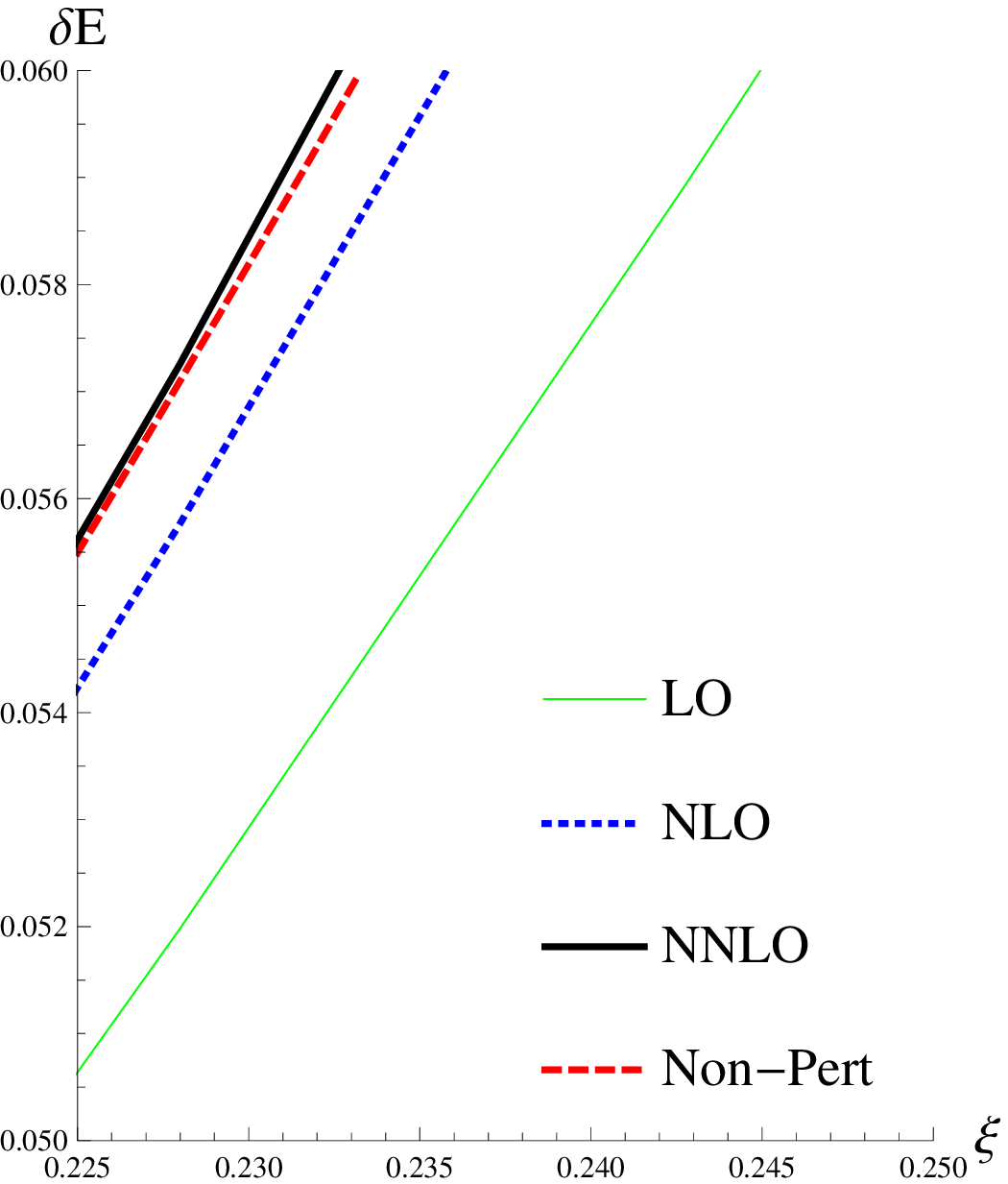,width=5.cm} 
\epsfig{file=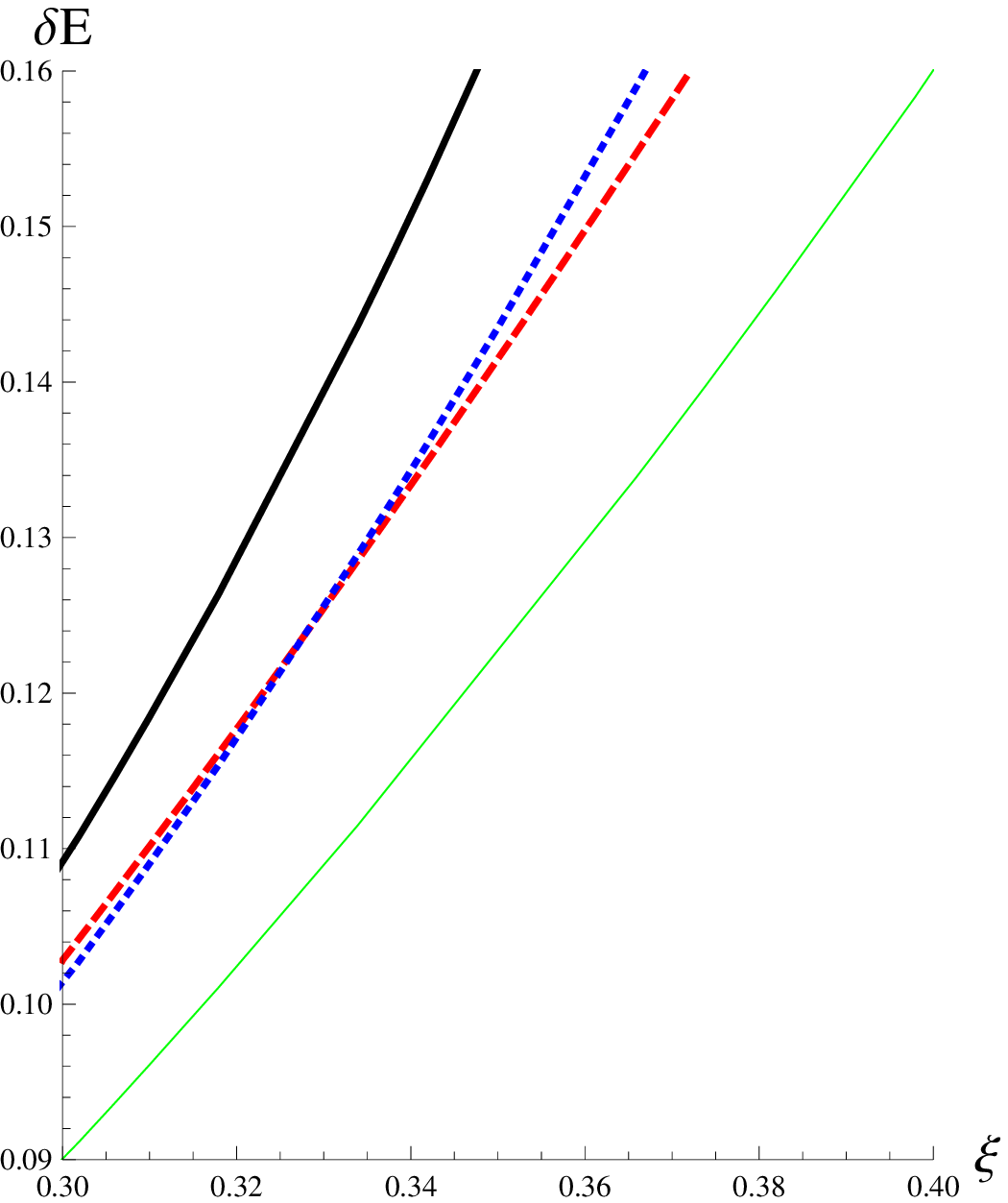,width=5.cm} 
\epsfig{file=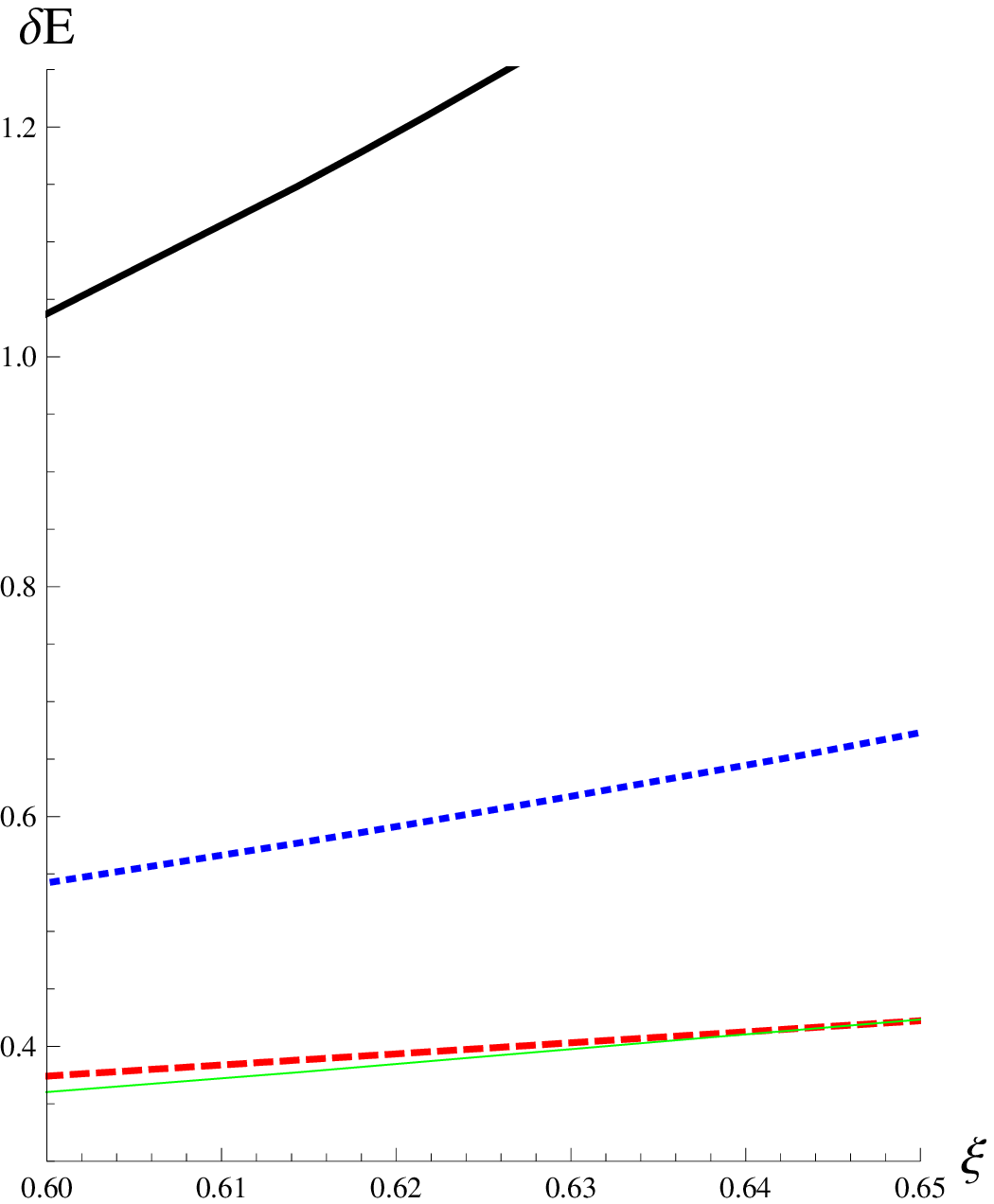,width=5.cm} 
        \caption{Comparison of perturbative and non-perturbative results in constant 
        electric fields. Plots are as in Figure~\ref{f:compare} with identical regions of $\xi$. 
        We have shown only the real part of the non-perturbative result.}
	\label{f:compareE}
\end{figure}
%

To investigate the non-perturbative result, we must perform
the analytic continuation. 
As shown in~\cite{Cohen:2007bt}, 
the correct analytic continuation is given by%
\footnote{
As a warm-up for the nucleon case, 
we derive an equivalent integral representation for the analytic
continuation of $i \cI(-i x)$ in Appendix~\ref{s:analcont}.
}
\begin{eqnarray}
i  \, \cI (-  i x )  
&=&
x \left( 1 - \gamma_E  - \log \frac{x}{2} \right)
- 
2  \tan^{-1} x
+ 
\sum_{n=1}^\infty
\left[ \frac{x}{n} - 2  \tan^{-1} 
\frac{x}{2n +1} 
\right]
- i  \log ( 1 + e^{- \pi x} ).
\notag \\
\label{eq:continue} 
\end{eqnarray}
With this expression we can evaluate the neutral pion 
energy in an electric field, Eq.~\eqref{eq:nonpertE},
and compare with the weak field expansion.  
This is done in Figure~\ref{f:compareE}, where we compare LO, NLO, and NNLO
results for the energy shift to the real part of the non-perturbative result. 
In each case, we plot $\d E$, in Eq.~\eqref{eq:deltaE},
for a constant electric field.  
Again we are working to a fixed order in the chiral expansion, 
and the expansion we are considering is in the strength of the external field.
We show identical ranges for the expansion 
parameter $\xi$ as for the magnetic case. 
Here, of course, we use $\xi = e |E| / \sqrt{6} m_\pi^2$.
In what we call the perturbative regime, the expansion parameter $\xi \sim 1/4$,
and the higher-order results better approximate the non-perturbative answer.
Because all terms in the perturbative expansion are now positive, 
it must be the case that the leading-order term (and some higher-order terms)
are less than the non-perturbative answer. 
In the transition regime, the expansion parameter $\xi \sim 1/3$,
and adding the NNLO and higher terms is no longer efficacious. 
This is where the NLO result makes its transit through the non-perturbative
answer. Once this happens, the addition of any higher-order terms
necessarily worsens the agreement. 
Increasing $\xi$ further, the LO result becomes as large as 
the non-perturbative answer. 
This last regime is typified for $\xi \sim 2/3$,
and is where 
all higher-order terms push the result 
away from the non-perturbative answer. 
Perturbation theory is of no avail; 
consequently the LO result agrees best. 
The agreement, moreover, is phenomenally good. 
%
\begin{figure}
\epsfig{file=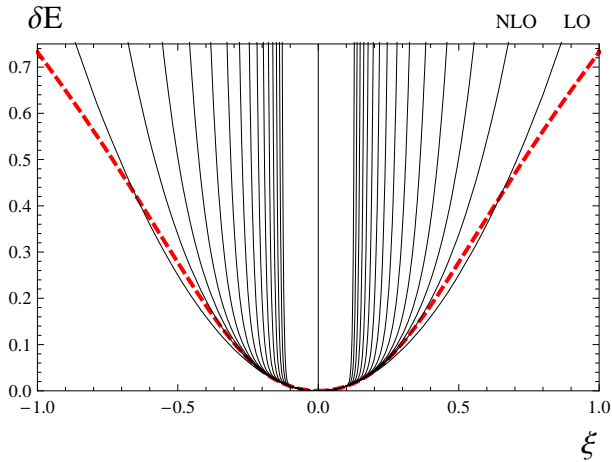,width=8cm} 
        \caption{Further comparison of perturbative and non-perturbative results
        in constant electric fields. The real part of the non-perturbative energy shift is plotted
        along with the first fifteen perturbative corrections in the strength of the field squared 
        (as in Figure~\ref{f:PvsNP}).
        }
	\label{f:PvsNPE}
	\end{figure}
%

In Figure~\ref{f:PvsNPE}, we compare perturbative
results up to fifteenth order in the expansion as we did above for the magnetic case. 
Because all terms in the perturbative expansion are positive, 
we see that for large enough $\xi$, all approximations overshoot
the non-perturbative result. 
Again we are only considering the real part of this result.
The higher-order approximations accumulate around $\xi=0$, as expected.
Beyond the perturbative regime, the LO result
gives a remarkably good approximation for 
$|\xi| \lesssim 1$.
Of course this approximation cannot be improved.

A feature of the non-perturbative result, is the imaginary 
part of the effective mass which we have so far neglected. 
This imaginary part has the form
\begin{equation}
\Im \mathfrak{m}
(
m^2_{ \text{eff}, \pi^0 }
)
= 
-
\frac{2 e |E| m_\pi^2}{( 4 \pi f)^2} 
\log 
\left[
1 
+
\exp
\left(
- \frac{\pi m_\pi^2}{e |E|}
\right)
\right]
,\end{equation}
which corresponds to a width
$\Gamma_{\pi^0}
= 
- 
\Im \mathfrak{m}
(
m^2_{ \text{eff}, \pi^0 }
) / m_\pi$.
The full width of the neutral pion is a sum of partial widths%
\footnote{
This decomposition follows from similar reasoning to that in the 
case of pair production from the vacuum, see, e.g.,~\cite{Nikishov:1969tt,Nikishov:1970br,Holstein:1999ta,Cohen:2008wz}. 
}
\begin{equation}
\Gamma_{\pi^0}
= 
\sum_{n=1}^\infty
\Gamma_n
,\end{equation}
where the partial widths correspond to decay modes with $n$-pairs
of charged pions
\begin{equation}
\Gamma_n  = \Gamma [\pi^0 \overset{E}{\longrightarrow} \pi^0 ( \pi^+ \pi^-)^n]
= 
(-1)^{n+1} 
\frac{ 2 e |E| m_\pi}{( 4 \pi f)^2}
\exp \left( - \frac{n \pi m_\pi^2}{e |E|} \right)
.\end{equation}
Virtual fluctuations in the pion can decay to real particles
in strong electric fields.

\subsection{A Connection with the Chiral Condensate}
\label{s:Condensate}

Now we demonstrate that the Gell-Mann--Oakes--Renner
relation is satisfied for the pions in electric and magnetic fields. 
The magnetic result for the neutral pion was previously 
demonstrated in~\cite{Shushpanov:1997sf}, 
however, that analysis was limited to the strict chiral 
limit, $m_\pi^2 = 0$. 
Here we keep the pion mass non-vanishing using the power counting
employed above, 
Eq.~\eqref{eq:PC},
and demonstrate the relation for both
charged and neutral pions.

In the absence of external electromagnetic fields, the 
Gell-Mann-Oakes-Renner relation~\cite{GellMann:1968rz}
reads
\begin{equation} \label{eq:GMORR}
f_\pi^2 m_\pi^2 
= 
4 < \ol q q > m
,\end{equation}
where $< \ol q q >$ is the quark condensate.
In the chiral limit, we have $f_\pi \to f$ and $< \ol q q > \to \lambda$
so that Eq.~\eqref{eq:GMOR} is recovered.
We ignore strong isospin breaking throughout so that
there is only one condensate:
$< \ol q q > = < \ol u u > = < \ol d d > $.  
Now imagine turning on a constant magnetic field, for example.
Above we have deduced the effective mass for charged and neutral pions. 
For the neutral pion, the effective mass is just the rest energy; while for the 
charged pion, this physical interpretation is not warranted because there 
are additionally Landau levels.
This technicality notwithstanding, we might imagine the following
generalization of Eq.~\eqref{eq:GMORR}
\begin{equation} \label{eq:GMORB}
f_\pi^2(B)  \, m_{\pi, \text{eff}}^2(B) 
= 
4 < \ol q q >_B  m
,\end{equation}
where we have displayed the dependence on the external field $B$. 
We can write each of the field-dependent quantities in terms
of their zero field value and the relative difference
\begin{eqnarray}
 m_{\pi, \text{eff}}^2(B) 
 &=& 
 m_\pi^2 \left[ 1  + \delta m_\pi^2(B) \right], \\
 f_\pi^2(B) 
 &=& 
 f_\pi^2 \left[ 1 + \d f_\pi^2(B) \right], \\
 < \ol q q >_B 
 &=& 
 < \ol q q >  \left[ 1 + \delta < \ol q q>_B \right]
.\end{eqnarray}
Taking the ratio of Eq.~\eqref{eq:GMORB} and Eq.~\eqref{eq:GMORR}, 
we have
\begin{equation} \label{eq:simple}
\d f_\pi^2(B) + \d m_\pi^2(B) = \d < \ol q q >_B
.\end{equation}
Because each of these differences arises at next-to-leading order 
$\sim \e^2$, it is obvious that terms second order in differences
can be dropped. 
To verify the conjectured relation Eq.~\eqref{eq:GMORB}, 
we shall merely evaluate
Eq.~\eqref{eq:simple}.
We have already determined the shift in the charged and neutral
pion effective mass as a function of the magnetic field $B$. 
The relative shift of the condensate is known:
from~\cite{Cohen:2007bt}, we have
\begin{equation}
\d < \ol q q >_B
=
- \frac{2 e |B|}{(4 \pi f)^2} \cI \left( \frac{m_\pi^2}{e |B|} \right) 
.\end{equation}
Hence to verify the low-energy theorem Eq.~\eqref{eq:GMORB}, we must check that
the change in pion decay constant compensates for the field dependence of the 
pion mass and chiral condensate:
\begin{eqnarray}
\label{eq:ToShow1}
\d f_{\pi^0}^2 (B) 
&=& 
-
\frac{4 e |B|}{(4 \pi f)^2} \cI \left( \frac{m_\pi^2}{e |B|} \right) , 
\\
\label{eq:ToShow2}
\d f_{\pi^{\pm}}^2 (B) 
&=& 
-
\frac{2 e |B|}{(4 \pi f)^2} \cI \left( \frac{m_\pi^2}{e |B|} \right) 
- 
\frac{8 e^2 B^2}{f^2 m_\pi^2}
( \a_9 + \a_{10} )
.\end{eqnarray}

For the neutral pion, calculation of the decay constant in a background
magnetic field is straightforward. The decay constant enters in
the pion-to-vacuum transition matrix element of the axial current
$\cA^a_\mu$
\begin{equation}
\int d^4 x \, e^{ - i p_\mu x_\mu} 
\langle 0 | \cA^a_\mu(x) | \pi^0 (0) \rangle
= \sqrt{2} \delta^{a 3} i p_\mu \,
f_{\pi^0} 
.\end{equation}
Because the neutral pion propagator maintains a simple pole form, 
we can apply LSZ to calculate this amplitude. 
At this order in the chiral expansion, there are no external field-dependent
local contributions to the axial current. This fact owes to the charge 
neutrality of the $\pi^0$, but does not continue to hold beyond next-to-leading order.
There are thus only two types of contributions to the external field dependence of 
$f_{\pi^0}$:
the wavefunction renormalization factor,
and the loop contribution from the next-to-leading order axial current.
The former has been determined in Eq.~\eqref{eq:pizeroZ}.
The next-to-leading order axial current
has the form
\begin{equation} \label{eq:ANLO}
\cA_{\mu}^a 
\overset{NLO}{=}
\frac{16 }{3 f} 
\left( D_\mu \pi^a \vec{\pi} \cdot \vec{\pi}  - \pi^a D_\mu \vec{\pi} \cdot \vec{\pi} \right)
.\end{equation}
When considering the $a=3$ current, 
recall that all the $D_\mu \phi$ terms must be contracted with the external pion,
else the loop contractions vanish. 
Adding up the loop contributions from the wavefunction correction
and next-to-leading order current, we arrive at
\begin{equation}
\frac{f_{\pi^0}(B)}{f_{\pi}}
=
\frac{f [ 1 - \frac{2}{f^2} \cD(B) ]}
{f [ 1 - \frac{2}{f^2} \cD(0)]}
= 
1 - \frac{2}{f^2} [ \cD(B) - \cD(0) ] 
+\ldots
,\end{equation}
where we have dropped higher-order terms. 
Squaring this result and using 
the definition of $\cI(x)$
yields the desired equality in Eq.~\eqref{eq:ToShow1}.

The charged pions, by contrast, cannot be handled in this manner. 
One problem with the approach is the amputation; and, 
instead we focus on the coordinate-space correlation function
$C_\mu^\pm(x)$, given by
\begin{equation}
C_\mu^\pm(x) 
=
\langle 0 | A_\mu^\pm(x)  \pi^\mp (0) | 0 \rangle
,\end{equation}
where it is assumed that $x_\mu \neq 0$ to avoid contact terms. 
At next-to-leading order in the chiral expansion, there are two 
possible Lorentz structures into which the correlator can be decomposed
\begin{equation} \label{eq:decomp}
C_\mu^\pm(x) 
=
- f_1(B) \,  D_\mu D(x,0)
+ 
2 i f_2(B) F_{\mu \nu} \, D_\nu D(x,0)
.\end{equation}
Terms with higher powers of the field strength tensor, 
such as $F_{\mu \nu} F_{\nu \rho} D_\rho$, 
must come from local terms in the effective theory, 
and are hence suppressed by at least 
two powers of the chiral symmetry 
breaking scale.
The field dependence of the coefficients $f_1(B)$ and $f_2(B)$
must be generated from charged pion loops. 
A quick power counting argument shows that $f_2(B)$ can only 
be generated from local interactions, hence this coefficient is
independent of the magnetic field, $f_2(B) = f_2$.

Taking the covariant four-divergence of the axial correlator, we
have
\begin{equation} \label{eq:axialcor}
D_\mu C_\mu^\pm(x) 
= 
- [ f_1(B) m_\pi^2 + f_2 Q B^2 ] D(x,0)
.\end{equation}
To arrive at this expression, we have used the fact that the propagator
is a Green's function Eq.~\eqref{eq:Green}, 
and the relation $2 F_{\mu \nu} D_\mu D_\nu = i Q F_{\mu \nu} F_{\mu \nu}$.
In this form, we can amputate the correlation function in coordinate space
\begin{eqnarray}
\langle 0 | D_\mu A_{\mu}^{\pm} | \pi^\pm(0) \rangle
&\equiv&
\int d^4 x
\left( - D_\mu D_\mu + m_\pi^2 \right) 
D_\mu C_\mu^\pm(x)  
.\end{eqnarray}
From this amplitude arises a natural generalization
of the pion decay constant
\begin{equation} \label{eq:chargedFpi}
\langle 0 | D_\mu A_{\mu}^{\pm} | \pi^\pm(0) \rangle
=
- m_\pi^2 f_{\pi^\pm} (B)
.\end{equation}
In the chiral limit, the axial current is 
conserved $\partial_\mu A^a_\mu(x) = 0$.
Thus the decay constant defined in 
Eq.~\eqref{eq:chargedFpi}
will be singular in the chiral limit
because the covariant divergence does not vanish. 
We are forced to use the covariant divergence, however, in 
order to utilize LSZ. 
Notice that the amplitude itself is finite in the chiral limit. 
The singularity in $f_{\pi^\pm}(B)$, moreover, must disappear when 
the external field is turned off.

Having generalized the defining relation for the pion decay constant, 
we can now calculate $f_{\pi^\pm}(B)$ to one-loop order. 
The wavefunction correction has been obtained above, Eq.~\eqref{eq:piplusZ}. 
Additionally one-loop diagrams arise from the 
next-to-leading order axial current in Eq.~\eqref{eq:ANLO}. 
Combining the results from loop diagrams, we can identify the 
factor $f_1(B)$ in Eq.~\eqref{eq:decomp}, 
namely
\begin{equation}
f_1(B) = f \left\{ 1 - \frac{1}{f^2}[ \cD(B) + \cD(0)]  \right\}
.\end{equation}

The coefficient $f_2$ arises from a local contribution to the axial 
current contained in the next-to-leading order Lagrangian.
The relevant external field dependent terms are
just those in Eq.~\eqref{eq:NLO}, but generalized
to include both left- and right-handed sources 
\begin{equation}
\cL 
=
- i \a_9
<  L_{\mu \nu} D_\mu \Sigma D_\nu \Sigma^\dagger 
+ 
 R_{\mu \nu}  D_\mu \Sigma^\dagger D_\nu \Sigma >
- 
\a_{10}
< 
L_{\mu \nu} \Sigma  R_{\mu \nu} \Sigma^\dagger 
>
\label{eq:NLOLR}
.\end{equation}
The left-handed field strength,
$L_{\mu \nu}$,
 is given by
$L_{\mu \nu} = \partial_\mu L_\nu - \partial_\nu L_\mu + i [ L_\mu, L_\nu]$,
and similarly for the right-handed field strength,
$R_{\mu \nu}$. 
The left- and right-handed four-vector potentials, 
$L_\mu$ 
and 
$R_\mu$,
are given by:
$L_\mu = Q e A_\mu + \tau^a \cA^a$, 
and
$R_\mu = Q e A_\mu - \tau^a \cA^a$.
The action of the chirally covariant derivative is specified by,
$D_\mu \Sigma = \partial_\mu \Sigma + i L_\mu \Sigma - i \Sigma R^\dagger$.
Evaluating the single pion terms of the axial current derived from Eq.~\eqref{eq:NLOLR}, 
we find the tree-level contribution 
\begin{equation}
\d \cA_\mu^\pm
= 
- \frac{1}{f} 8 i ( \a_9 + \a_{10} ) Q e F_{\mu \nu} D_\nu \pi^{\pm}
,\end{equation}
with $\d \cA_\mu^3 = 0$. 
From this expression, we can calculate the axial correlator and identify $f_2$ in Eq.~\eqref{eq:decomp}.
We find the value $f_2 = - 4 ( \a_9 + \a_{10} ) Q e / f$. 
Combining the tree-level and loop contributions, we deduce
the charged pion decay constant using Eq.~\eqref{eq:chargedFpi}
\begin{equation}
\frac{f_{\pi^\pm}(B)}{f_\pi}
=
1 
- \frac{1}{f^2} [ \cD(B) - \cD(0) ] 
- \frac{4  e^2 B^2}{f^2 m_\pi^2} (\a_9 + \a_{10} )
.\end{equation}
Upon squaring, we find that Eq.~\eqref{eq:ToShow2} is indeed satisfied.

We have shown that the low-energy theorem for the neutral pion in a magnetic
field~\cite{Shushpanov:1997sf} extends beyond the chiral limit into the 
regime where $m_\pi^2 \sim e |B|$. Furthermore we have generalized the
definition of the charged pion decay constant for a background field. 
This generalization enabled us to demonstrate that the Gell-Mann--Oakes--Renner
relation holds additionally for charged pions in magnetic fields, Eq.~\eqref{eq:GMORB}. 
There is no subtlety in repeating the calculation for the case of a
Euclidean electric field. The effective mass was deduced above in 
Section~\ref{s:PiE}. The chiral condensate is Lorentz invariant, and the determination
of the decay constant was Lorentz covariant. Thus we arrive at
the $\cE$-field results for $< \ol q q >_\cE$ and $f_\pi(\cE)$ merely by substituting $B \to \cE$. 
The final step to deriving the low-energy theorem in an electric field
is the analytic continuation. 
The tree-level contributions are trivial to continue, while the one-loop
expression requires Eq.~\eqref{eq:continue}. 
After analytic continuation, we arrive at
\begin{equation} \label{eq:GMORE}
f_\pi^2(E)  \, m_{\pi, \text{eff}}^2(E) 
= 
4 < \ol q q >_E  m
,\end{equation}
for $\pi = \pi^0$, and   $\pi^\pm$.

%
\section{Nucleon in Strong Fields}
\label{s:Nuc}

\subsection{Nucleon Chiral Lagrangian}
\label{s:NucL}

To describe the physics of nucleons in strong electric 
and magnetic fields, 
we use the nucleon chiral Lagrangian. 
A well known complication arises in trying
to describe the chiral dynamics of the nucleon. 
In the chiral limit, the nucleon has a non-vanishing mass, 
$M_N$,
which is on the order of the chiral symmetry breaking scale,
$M_N \sim \Lambda_\chi$. 
Consequently there is no power counting to order
the tower of operators needed to renormalize the 
theory of nucleons and pions, e.g. terms with any number of derivatives
acting on the nucleon field are $\cO(\e^0)$. 
The solution to this complication is to treat the nucleon
non-relativistically. 
A particularly clean implementation of the non-relativistic 
expansion for nucleons is known as heavy baryon 
chiral perturbation theory~\cite{Jenkins:1990jv,Jenkins:1991es},
which borrowed earlier developments from heavy quark effective 
theory~\cite{Georgi:1990um,Manohar:2000dt}.

Consider a fermion field $\cB$ of mass $M_\cB$, with $M_\cB = M_N+\D$. 
We choose this consideration because we will 
include both nucleons (for which we can choose $\D = 0$), and deltas
(for which consequently $\D \neq 0$) into chiral perturbation theory. 
The free Lagrangian for $\cB$ in Euclidean space appears as
\begin{equation} \label{eq:FreeB}
\cL = \ol \cB \left( \rlap \slash \partial +  M_\cB \right) \cB
.\end{equation}
The mass terms spoils the power counting when pion interactions
are included. 
To remove this term, we expand about the non-relativistic solution 
to the classical equations of motion
\begin{equation} \label{eq:redef}
\cB_v(x) = \cP_+ e^{- i M_N v_\mu x_\mu} \cB(x)
.\end{equation}
Here 
$v_\mu$
is the covariant four-velocity. In the rest frame,
$v_\mu = (0,0,0,i)$, 
so that the velocity satisfies the Euclidean space
relation 
$v_\mu v_\mu = -1$.  
The conjugate field 
$\ol \cB_v$ 
satisfies the conjugate relation without 
$v_\mu$ 
complex conjugated. In Euclidean space, 
we are free to treat the field and its conjugate as independent
variables. 
The projector 
$\cP_+$ 
is given in an arbitrary frame by 
$\cP_+ = \frac{1}{2} ( 1 - i  \rlap \slash v )$. 
Inserting the field redefinition, Eq.~\eqref{eq:redef}, into the free Lagrangian, Eq.~\eqref{eq:FreeB}, 
we find
\begin{equation}
\cL = \ol \cB_v ( - i v_\mu \partial_\mu  + \Delta ) \cB_v
.\end{equation}
Having phased away the large mass $M_N$, we are free to treat the residual mass 
$\Delta$ as a small parameter.
Additionally with the Fourier modes, $p_\mu$, of the original $\cB$ field 
written as  
$p_\mu = M v_\mu + k_\mu$,
we see the residual momentum $k_\mu$ 
is produced by derivatives acting on $\cB_v$. 
For slowly moving baryons, all components of $k_\mu$ are necessarily small.

Before writing down the heavy nucleon chiral Lagrangian, 
let us derive the free propagator for the $\cB$ field in the static limit. 
The heavy baryon two-point function, $D_\cB(x,0)$, has the form
\begin{eqnarray}
D_\cB(x,0) 
&\equiv&
\langle 0 | \cB(x) \ol \cB(0) | 0 \rangle
= 
 e^{i M_N v_\mu x_\mu} \langle 0 | \cP_+ \cB_v(x) \ol \cB_v(0) \cP_+ | 0 \rangle 
\notag \\
&=&
-e^{ - M_N x_4}
\int \frac{d^4 k}{(2 \pi)^4}
\frac{\cP_+ e^{- i k_\mu x_\mu}}{k_\mu v_\mu - \D}
=
\cP_+ \delta(\bm{x}) \theta(x_4) e^{- ( M_N + \D) x_4}
\label{eq:nukeprop}
,\end{eqnarray} 
where, after the first line, we have used the rest frame.
This propagator is the essential new ingredient needed
to perform chiral loop calculations for the nucleon.

To describe the nucleon, we use a two component isospinor $N_i$, 
given by $N = ( p, n )^T$. 
Additionally we include the nearest lying resonances, the deltas. 
These give rise to generally important virtual corrections to nucleon
observables in chiral dynamics, because the mass splitting 
$\D$ between the nucleons and deltas is on the order of the pion mass. 
Furthermore the pion-nucleon-delta axial coupling is not small. 
The deltas we describe using a completely symmetric flavor tensor 
$T_{ijk}$, 
where: 
$T_{111} = \D^{++}$, 
$T_{112} = \D^+ / \sqrt{3}$,
$T_{122} = \D^0 / \sqrt{3}$, 
$T_{222}= \D^-$, 
\emph{etc}.
To keep a consistent power counting, we must expand the
$N$ and $T$ 
fields about their classical equations of motion.
We shall leave the velocity subscript as implicit.  
Because the nucleons and deltas interact, 
we cannot phase away both chiral limit masses 
$M_N$ and $M_T$. 
Instead we phase away 
$M_N$ 
from both multiplets. This leaves a 
residual mass 
$\D = M_T - M_N$ 
for the deltas.
Thus we have the following additions to the power counting in 
Eq.~\eqref{eq:PC}
\begin{equation} \label{eq:PCnew}
\frac{k_\mu}{M_N} 
\sim
\frac{\D}{M_N}
\sim \e
.\end{equation}
We will not need to distinguish between the heavy baryon 
mass, and the chiral symmetry breaking 
scale, i.e.~we treat 
$M_N \sim \L_\chi$.

The relevant terms of the heavy nucleon and delta chiral Lagrangian at 
$\cO(\e)$ 
are
\begin{equation} \label{eq:Nuke}
\cL = \ol N (-  i v_\mu \cD_\mu ) N
+ 
\ol T_\nu (-  i v_\mu \cD_\mu + \D) T_\nu
+ 2 g_A  \ol N S_\mu \cA_\mu  N 
+ 2 g_{\D N} [ ( \ol T_\mu \cA_\mu N) + ( \ol N \cA_\mu T_\mu ) ]
.\end{equation}
The vector $S_\mu$ is the covariant spin operator satisfying
$S_\mu S_\mu = 3/4$. 
Electromagnetism is coupled to the Largrangian through the axial 
$\cA_\mu$ 
and vector 
$\cV_\mu$ 
fields of pions given by
\begin{eqnarray}
\cV_\mu &=& i e A_\mu \cQ +  \frac{1}{2f^2} \left(  \phi D_\mu \phi - D_\mu \phi \, \phi \right) \ldots, 
\\
\cA_\mu &=& \frac{1}{f} D_\mu \phi   + \ldots,
\end{eqnarray}
where higher-order terms have been dropped.
The chirally covariant derivative
$\cD_\mu$
has the action
\begin{equation}
\left( \cD_\mu N \right)_i 
=
\partial_\mu N_i 
+ 
\left( \cV_\mu \right)_i {}^{i'} \, N_{i'}
+
\tr \left( \cV_\mu \right) \, N_i
,\end{equation}
on the nucleon field, and
\begin{equation}
\left( \cD_\mu T \right)_{ijk}
=
\partial_\mu T_{ijk}
+
\left( \cV_\mu \right)_i {}^{i'} T_{i'jk}
+
\left( \cV_\mu \right)_i {}^{j'} T_{ij'k}
+
\left( \cV_\mu \right)_i {}^{k'} T_{ijk'}
,\end{equation}
on the delta.

\subsection{Nucleon in Strong Magnetic Fields}
\label{s:NucB}

Having outlined the chiral Lagrangian for the nucleon, 
we now determine the energy of a nucleon 
in an external magnetic field to 
$\cO(\e^2)$
by computing corrections to the nucleon two-point function.
Corrections to the nucleon's energy
in an external field arise from pion loops. 
As in the meson sector, 
these contributions are non-perturbative in the field strength,
$e |B| / m_\pi^2 \sim \e^0$. 
There are additionally local interactions, 
but the coefficients of these terms are 
phenomenologically known. 
Thus we will be able to make quantitative 
predictions for nucleons in strong fields.

In calculating the nucleon two-point function to $\cO(\e^2)$, 
there are both local contributions and loop contributions.
In the effective field theory, 
the local interactions arise from magnetic moment operators.
For the nucleons, there are isovector and isoscalar operators
having the form~~\cite{Jenkins:1992pi,Bernard:1992qa}
\begin{equation} \label{eq:nukemu}
\cL 
= 
- \frac{i e}{2 M_N} F_{\mu \nu} 
\left( 
\mu_0 \ol N [ S_\mu, S_\nu ] N
+
\mu_1 \ol N [ S_\mu, S_\nu ] \tau^3 N
\right)
.\end{equation}
There is only one additional operator needed to this order:
the nucleon-delta magnetic dipole transition operator~\cite{Jenkins:1992pi}
\begin{equation}
\cL 
= 
- 3 \sqrt{3}\, 
\mu_T 
\frac{i e}{2 M_N} 
F_{\mu \nu}
\left[
\left( \ol N S_\mu \cQ T_\nu \right)
+
\left( \ol T_\mu \cQ S_\nu N \right)
\right].
\end{equation}
In the absence of other corrections, these terms lead to 
a shift in the nucleon energy of the form
\begin{equation}
E_N(B) = M_N - \mu_N \frac{e B \sigma_3}{2 M_N} -  \frac{\mu_T^2}{\D} \left( \frac{e B}{2 M_N} \right)^2 
.\end{equation}
The magnetic moments denoted by $\mu_N$ are isospin dependent, 
$\mu_p = \mu_0 + \mu_1$, 
and 
$\mu_n = \mu_0 - \mu_1$. 
The nucleon magnetic moment term scales as 
a relative $\cO( \e)$ correction to the mass $M_N$,
and the magnetic transition term scales as $\cO ( \e^2)$. 
The isovector magnetic moment $\mu_1$
also serves as a counterterm to renormalize
the $\cO(\e^2)$ loop contributions.
Technically one needs additional operators
having exactly the same structure as those in 
Eq.~\eqref{eq:nukemu}, 
but multiplied by the chiral singlet parameter 
$\D / M_N$.  
In addition to renormalizing contributions to the nucleon mass, 
we will additionally renormalize the magnetic moment contributions
so that 
$\mu_1$ 
takes its physical value.

%
\begin{figure}
\epsfig{file=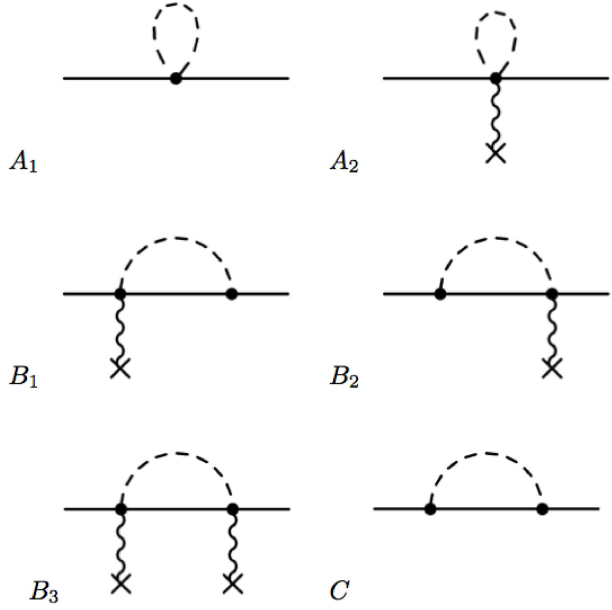,width=8cm} 
        \caption{Diagrams for the nucleon two-point function in an external field.
          The straight lines denote baryons, while the dashed lines represent the background
        field propagator of the pion. Wiggly lines terminating in crosses
        represent couplings of the interaction vertices to the background field.
        The internal baryons lines can be both nucleons and deltas.
        }%
	\label{f:nucleon}
	\end{figure}
%

There is one further local operator that leads to a nucleon energy shift in an external magnetic field. 
This correction has a coefficient exactly fixed by gauge invariance and reparametrization 
invariance~\cite{Luke:1992cs}. The latter arises as a convenient way to impose Lorentz
invariance order-by-order in the heavy baryon expansion. 
Applying reparametrization invariance on the leading kinetic term for the nucleon,
we generate the kinetic energy operator
\begin{equation}
\cL =   - \ol N \frac{\cD_\perp^2}{2 M_N} N
,\end{equation}
where $(\cD_\perp)_\mu = \cD_\mu + v_\mu ( v_\nu \cD_\nu)$.
This operator leads to corrections to the nucleon energy at
$\cO(\e^2)$, thus we do not need the chiral 
structure in the derivative $\cD_\mu$. Furthermore, 
for the neutron, we completely eliminate this contribution
by keeping the neutron at rest. 
For the proton, however, the gauged kinetic energy leads to Landau levels.
For zero transverse momentum, the proton Landau levels are given by
\begin{equation}
E_p(n) =  \frac{e |B|}{M_N}  \left( n + \frac{1}{2} \right) 
,\end{equation}
where $n$ is an integer. 
Although these energies scale as $\cO(\e)$, 
large values of $n$ will upset the power counting, 
so we are necessarily restricted to only the lowest levels. 
Relativistic corrections to this result give rise to an energy shift
at $\cO(\e^3)$ in the power counting.

The remaining contributions to the nucleon energy are non-analytic in nature,
and stem from pion loop diagrams generated from chiral dynamics.
The interactions between baryons and pions are contained in Eq.~\eqref{eq:Nuke}. 
Using the vertices appearing in this Lagrangian, we generate the 
diagrams shown in Figure~\ref{f:nucleon} for the nucleon two-point function. 
Most importantly there are no charge couplings of the photon to the nucleon or delta. 
These have been eliminated from Eq.~\eqref{eq:Nuke} by the gauge condition
$v_\mu A_\mu =0$. Consequently the background field only couples to the pions,
or to the interaction vertices.

Of the diagrams shown in the figure, five are trivial to evaluate. 
Diagrams $A_1$ and $A_2$ arise from expanding out the vector 
field of pions $\cV_\mu$ to include two pion terms. 
The partial derivative piece generates diagram $A_1$, 
while the gauge piece of the derivative generates $A_2$. 
Both of these diagrams vanish because the vector field is 
dotted into the four-velocity $v_\mu$. 
In the case of $A_1$, the derivative acting 
inside the loop picks off a good component of momentum,
making the momentum integrand odd; in the case of $A_2$,
we have $v_\mu A_\mu = 0$.
While the remaining diagrams $B_1$--$B_3$, and $C$ are all 
connected by gauge invariance, 
we find it simpler to evaluate each separately. 
A useful simplification to note stems from the Dirac-delta function
in the heavy nucleon propagator, Eq.~\eqref{eq:nukeprop}.
We can choose the nucleon source to be located at
position $\bm{x} = \bm{0}$. Because the nucleon remains static, 
the Dirac-delta function enforces the intermediate and final states
also to be at $\bm{x} = \bm{0}$. 
For diagrams $B_1$--$B_3$ coupling of the external field
to the vertices introduces an explicit power of the potential 
$A_\mu$, which depends on the spatial coordinates. 
These diagrams thus vanish upon imposing 
the Dirac-delta function on position.
The only non-vanishing loop contribution to the nucleon 
two-point function in an external magnetic field is from 
diagram $C$. 
As the evaluation is rather technical, we relegate
the calculation and subsequent renormalization 
to Appendix~\ref{s:nukeBcalc}.

The complete expression for the nucleon energy in a background
magnetic field up to $\cO(\e^2)$ is given by
\begin{eqnarray}
E_N(B) 
= 
M_N + \frac{Q_N e |B|}{ 2 M_N}  
- 
e B \sigma_3 
\left( 
\frac{ \mu_N}{2 M_N} 
+
\tau_3 \, 
\d E_1(B^2)
\right)
- 
\frac{\mu_T^2}{\D} \left( \frac{e B}{2 M_N} \right)^2
+ 
\d E_2(B^2).
\notag \\ \label{eq:nukeEinB}
\end{eqnarray}
Here $\tau_3$ is the usual isospin matrix, $\tau_3 = \diag ( 1, -1)$, 
and we have specialized to the case of the lowest Landau level. 
The non-analytic functions 
$\d E_1(B^2)$ and $\d E_2(B^2)$ 
arise from pion loop graphs, and are given by
\begin{eqnarray}
\d E_1(B^2)
&=&
\frac{  \sqrt{\pi}}{  (4 \pi f)^2 }
\int_0^\infty \frac{ds}{s^{3/2}} e^{ - s m_\pi^2}
\left( 
\frac{2 e B s}{ \sinh 2 e B s} 
- 
1
\right)
\left[
g_A^2 +
\frac{2}{9} g_{\D N}^2 
\, e^{ s \D^2} 
\Erfc \left( \sqrt{s} \D \right) 
\right],
\notag \\
\label{eq:E1B}
\\
\d E_2(B^2)
&=&
\frac{m_\pi^2 \sqrt{\pi}}{4 ( 4 \pi f)^2}
\int_0^\infty \frac{ds}{s^{ 3/2}} 
e^{- s m_\pi^2}
\left( 
\frac{2 e B s}{\sinh 2 e B s}  - 1
\right)
\Bigg\{
g_A^2
+
\frac{8}{9} 
g_{\D N}^2
\notag \\
&& \phantom{spacersesingseseses}
\times 
\left.
\left[
\frac{1}{\sqrt{\pi s}} 
\frac{\D}{m_\pi^2}
+ 
\left( 
1 - \frac{\D^2}{m_\pi^2} 
\right)
e^{ s \D^2}
\Erfc \left( \sqrt{s} \D \right)
\right]
\right\}
\label{eq:E2B}
.\end{eqnarray} 
Notice we have renormalized the function $\d E_1(B^2)$ so that $\d E_1 (0) = 0$. 
In this renormalization scheme, $\mu_N$ is the physical nucleon magnetic moment.
Thus we have absorbed the well-known one-loop chiral corrections to magnetic moments
that are linear in $m_\pi$ into the coefficient $\mu_N$. 
This is possible because we are considering the external field dependence and
not the pion mass dependence of the nucleon energy.

In weak fields, we can perturbatively expand 
these results to arrive at the asymptotic series
\begin{eqnarray}
\d E_1(B^2) &=& \sum_{n=1}^\infty \d E_1^{(2 n)} B^{2n} ,
\qquad \text{and} \qquad
\d E_2(B^2) = \sum_{n=1}^\infty \d E_2^{(2 n)} B^{2n} 
.\end{eqnarray}
The first term in the expansion of $\d E_1(B^2)$, which gives a contribution to
the nucleon energy at $\cO(B^3)$, is
\begin{eqnarray}
\d E_1^{(2)} 
= 
\frac{-e^2}{48 \pi f^2}
\left\{
\frac{g_A^2}{m_\pi^3}
+ 
\frac{2 g_{\D N}^2}{9 \pi}
\left[
\frac{2 \D}{m_\pi^2 ( \D^2 - m_\pi^2)}
-
\frac{1}{(\D^2 - m_\pi^2)^{3/2}}
\log \left(
\frac{\D - \sqrt{\D^2 - m_\pi^2}}{\D + \sqrt{\D^2 - m_\pi^2}} 
\right)
\right]
\right\} .
\notag \\
\label{eq:E12}
\end{eqnarray}
In the chiral limit, this coefficient is singular and consequently 
highly constrained by the  parameters in the chiral Lagrangian. 
In general, the chiral limit behavior of 
terms in the asymptotic expansion are
\begin{equation}
\d E_1^{(2 n)} \sim m_\pi^{-4n + 1}, 
\quad \text{and}
\quad
\d E_2^{(2n)} \sim m_\pi^{-4n + 3}
.\end{equation}
The first term in the expansion of $\d E_2(B^2)$ gives a contribution to the
nucleon energy at $\cO(B^2)$ of the form
\begin{eqnarray} 
\d E_2^{(2)}
&=& 
- \frac{e^2}{192 \pi f^2}
\left[
\frac{g_A^2}{m_\pi}
-
\frac{8 g_{\D N}^2}{9 \pi} 
\frac{1}{\sqrt{\D^2 - m_\pi^2}}
\log \left(
\frac{\D - \sqrt{\D^2 - m_\pi^2}}{\D + \sqrt{\D^2 - m_\pi^2}} 
\right)
\right] \label{eq:E22}
.\end{eqnarray}
This energy shift is the chiral loop contribution 
to the nucleon magnetic polarizability, namely 
$\beta_M^\text{loop} = - \frac{1}{2 \pi} \d E_2^{(2)}$.  
The remaining contribution to $\b_M$ comes from the 
transition moment term proportional to $\mu_T^2$ in Eq.~\eqref{eq:nukeEinB}.
The derived value for $\b_M$ agrees with the known chiral 
perturbation theory expression~\cite{Bernard:1991rq,Bernard:1991ru,Hemmert:1996rw}.
The next correction in the expansion of $\d E_2(B^2)$, 
gives a shift at $\cO(B^4)$ to the
nucleon energy of the form
\begin{eqnarray}
\d E_2^{(4)}
&=&
\frac{7 e^4}{768 \pi f^2}
\left\{
\frac{g_A^2}{m_\pi^5}
+
\frac{32  g_{\D N}^2}{27 \pi} \frac{1}{(\D^2 - m_\pi^2)^2}
\left[
\frac{\D^3}{m_\pi^4}
- \frac{5 \D}{ 2 m_\pi^2}
\right. \right.
\notag \\
&& \phantom{spacingspacerspace}
\left. \left. 
- 
\frac{3}{4 \sqrt{\D^2 - m_\pi^2}}
\log \left(
\frac{\D - \sqrt{\D^2 - m_\pi^2}}{\D + \sqrt{\D^2 - m_\pi^2}} 
\right)
\right]
\right\}
\label{eq:E24}
.\end{eqnarray}

%
\begin{figure}
\epsfig{file=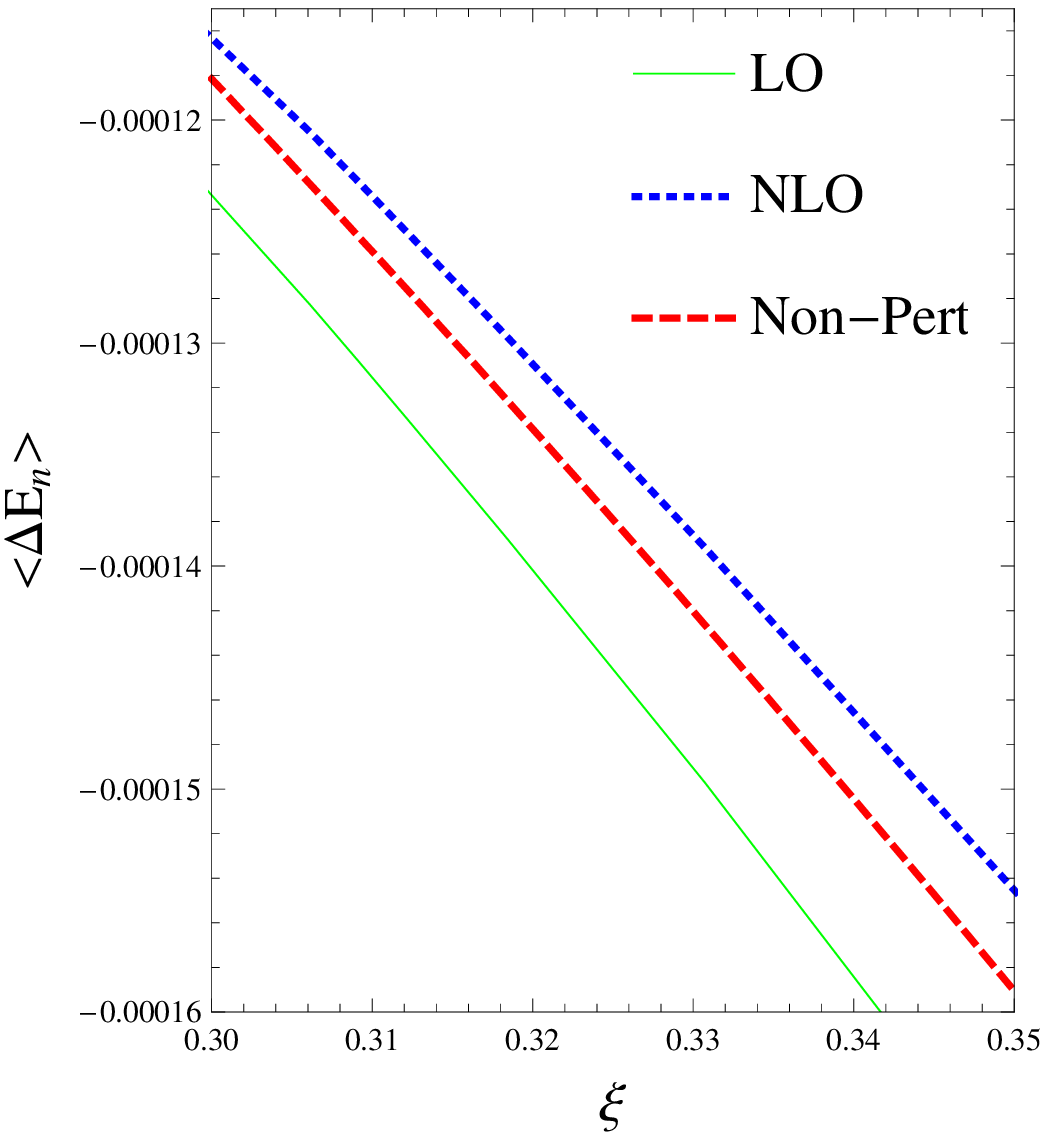,width=5.75cm} 
\epsfig{file=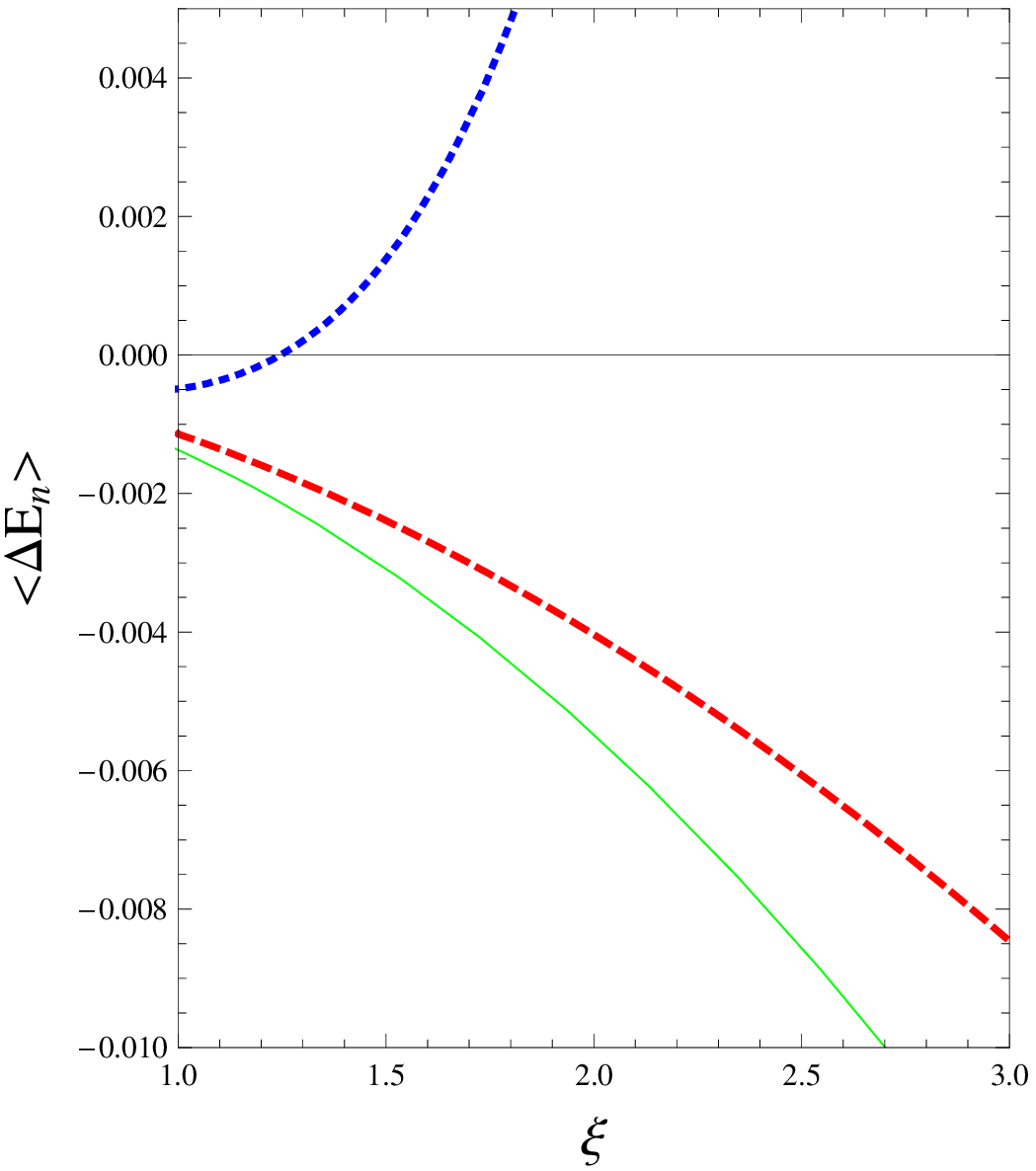,width=5.5cm} 
        \caption{Comparison of perturbative and non-perturbative results for relative shift in unpolarized neutron energy in constant 
        magnetic fields. Plotted is the non-perturbative result for $< \D E_n >$ in Eq.~\eqref{eq:neutronE} 
        as well as the LO and NLO weak-field approximations to this result. The expansion parameter $\xi = e |B| / m_\pi^2$.
        }
	\label{f:nukeB2}
	\end{figure}
%

From these first few terms in the weak-field series expansion, 
we can compare the asymptotic expansion
with the non-perturbative energy shift.
The energy of an unpolarized neutron
has an expansion in only even powers of the magnetic field. 
We consider the relative shift in the unpolarized neutron energy 
\begin{equation} \label{eq:neutronE}
< \D E_n > 
= 
- 
\frac{\mu_T^2}{\D M_N} \left( \frac{e B}{2 M_N} \right)^2
+ 
\frac{\d E_2(B^2)}{M_N}.
\end{equation}
The LO expansion of $<\D E_n>$ is due to the polarizability 
[from the transition moment and Eq.~\eqref{eq:E22}], 
while at NLO we add to the polarizability
the fourth-order term in Eq.~\eqref{eq:E24}. 
We can assess the behavior of the asymptotic series
using phenomenological input for the low-energy constants.
For the nucleon-delta mass splitting, we take 
$\D = 0.29 \, \texttt{GeV}$. 
The nucleon axial couplings is well known from experiment, 
$g_A = 1.25$.
The nucleon-delta axial coupling, $g_{\D N} = 1.5$,
we estimate from the width $\Gamma(\D\to N \pi)$~\cite{Amsler:2008zz}. 
The dipole transition coefficient is also known from $\D \to N \gamma$, 
giving the value
$\mu_T = 1.4$~\cite{Butler:1993ht,Gellas:1998wx,Arndt:2003vd}. 
In Figure~\ref{f:nukeB2}, we plot the non-perturbative shift 
$< \D E_n >$
along with the LO and NLO perturbative results.
The expansion parameter $\xi$ is taken to be $\xi = e |B| / m_\pi^2$. 
The two graphs contrast the perturbative regime, $\xi \ll 1$, where the higher-order correction
gives better agreement with the non-perturbative result, and the non-perturbative 
regime, $\xi \gtrsim 1$, where LO is the best available approximation but cannot be
improved.

\begin{figure}
\epsfig{file=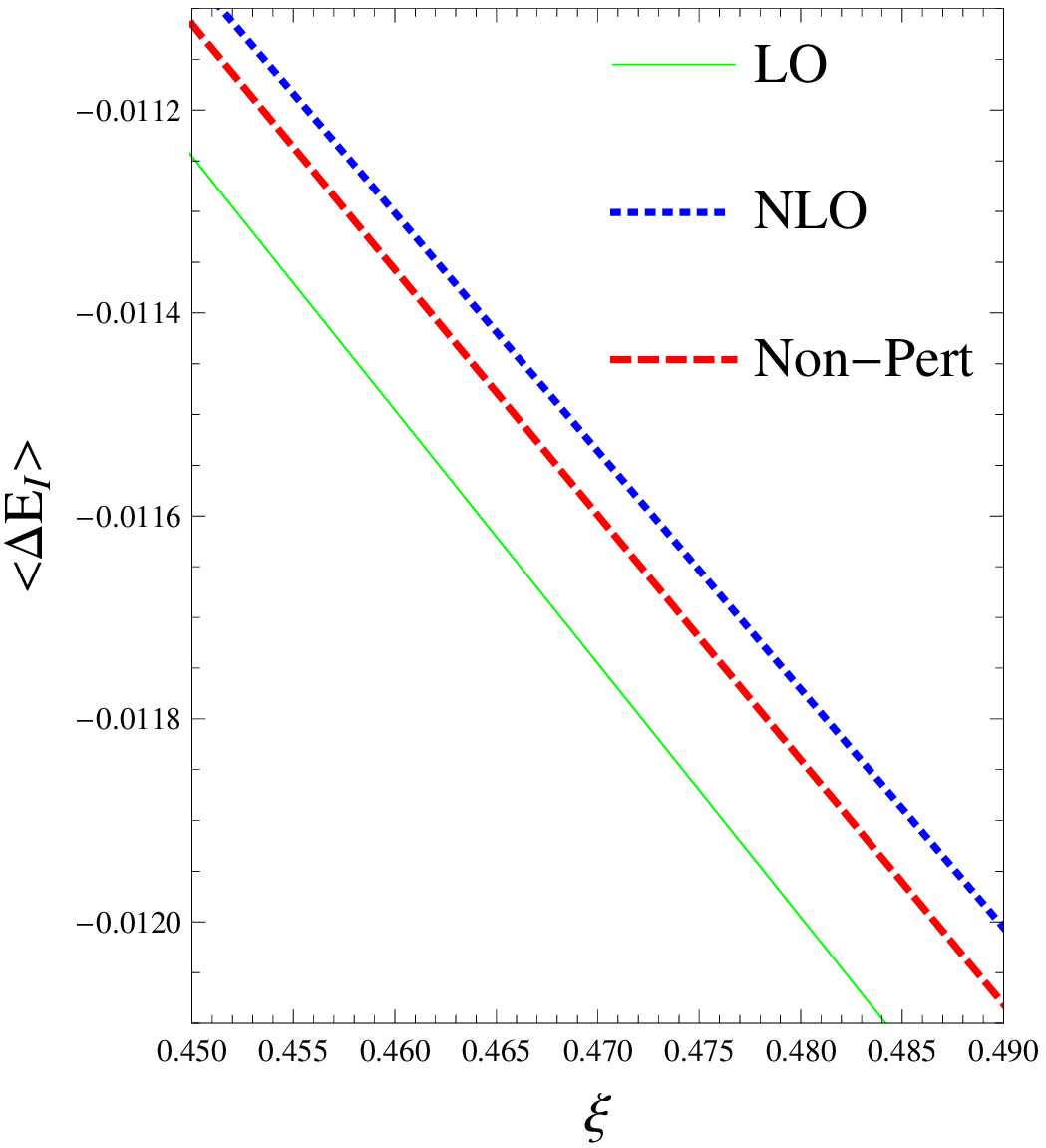,width=5.75cm} 
\epsfig{file=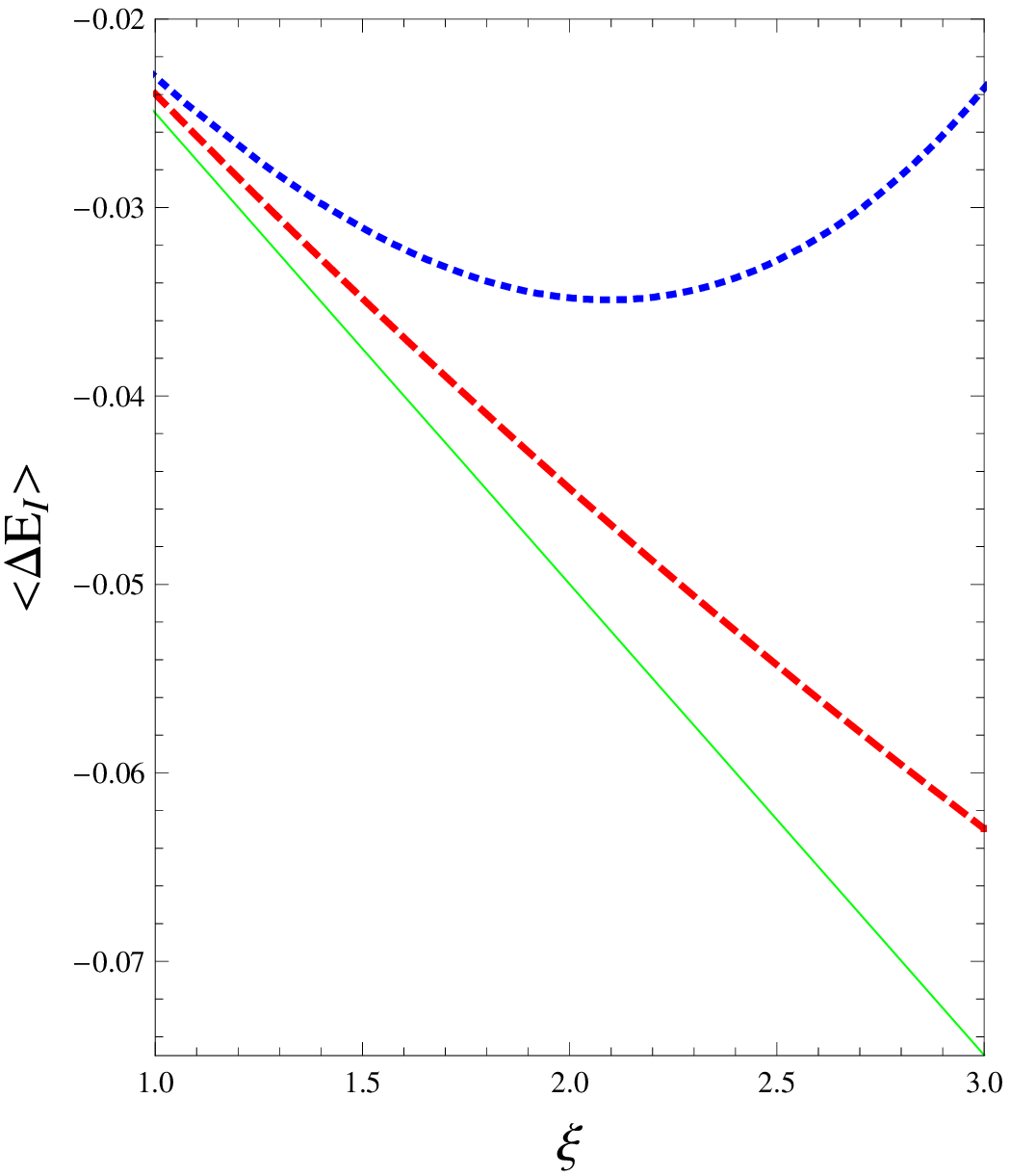,width=5.5cm} 
        \caption{Comparison of perturbative and non-perturbative results for the relative shift in the spin averaged, isospin averaged   
         nucleon energy in constant 
        magnetic fields. Plotted is the non-perturbative result for $< \D E_I>$ in Eq.~\eqref{eq:isovectorE} 
        as well as the LO and NLO weak-field approximations to this result. The expansion parameter $\xi = e |B| / m_\pi^2$.
        }
	\label{f:nukeB1}
	\end{figure}

The polarized isovector matrix element of the nucleon
has an expansion in only odd powers of the magnetic field. 
Taking the isospin average, and average of spin-up minus spin-down 
contributions, we find the relative shift in the energy
\begin{equation} \label{eq:isovectorE}
< \D E_I > 
= 
- 
e B 
\left( 
\frac{ \mu_1}{2 M^2_N} 
+
\frac{\d E_1(B^2)}{M_N}
\right)
.\end{equation}
The LO term in the expansion is just the linear dependence on the magnetic
field due to the isovector magnetic moment, $\mu_1 = 2.35$. 
The NLO term adds on the $B^2$ correction given in Eq.~\eqref{eq:E12}.
In Figure~\ref{f:nukeB1}, we plot the non-perturbative shift 
$< \D E_I >$
along with the LO and NLO perturbative expansions.
The expansion parameter $\xi$ is again taken to be $\xi = e |B| / m_\pi^2$, 
and we depict the behavior of $<\D E_I>$ in the perturbative 
and non-perturbative regimes.

%
%

The proton-neutron splitting is 
an interesting quantity to consider
in strong magnetic fields.
Taking the difference in ground state energies
of the proton and neutron,  we have 
\begin{eqnarray}
\Delta E_{pn}
&\equiv&
E_p(B) - E_n(B)
\notag \\
&=&
M_p - M_n
-
\frac{e |B|}{ M_N}  
\left( \mu_0 - \frac{1}{2} \right)
.\end{eqnarray}
Notice that both loop contributions, $\d E_1(B^2)$ and $\d E_2(B^2)$, cancel out
leaving the estimate of~\cite{Bander:1992ku}. 
The first chiral loop contributions to $\D E$ occur from the spin-dependent
isoscalar terms, and spin-independent isovector terms both present
in the $\cO(\e^3)$ expression for the energy, whereas
here we work to $\cO(\e^2)$.  
Using the value for the isoscalar magnetic moment, $\mu_0 = 0.44$, 
we find that the nucleon is stable to weak decay,
$\D E_{pn} = 0$,
for
$\xi = 1.0$.
This corresponds to 
a magnetic field strength of  
$B = 3.3 \times 10^{14} \, \texttt{T}$. 
Proton beta decay to neutron, positron, and neutrino
occurs when $\D E_{pn} \approx m_e$, 
or when 
$\xi = 1.5$
corresponding to 
$B = 4.9 \times 10^{14} \, \texttt{T}$.  
It is interesting to note that these values of the  field strength are exactly
in the regime where the power counting in 
Eqs.~\eqref{eq:PC} and \eqref{eq:PCnew}
applies. 
Corrections to these estimates from $\cO(\e^3)$ terms 
are proportional to $m_\pi / M_N \sim 15 \%$.

\subsection{Nucleon in Strong Electric Fields}
\label{s:NucE}

We now determine the one-loop
corrections to the nucleon two-point function
in an external electric field. 
We begin first by considering the Euclidean 
space electric field specified by the 
gauge potential in Eq.~\eqref{eq:AE}. 
After deriving the two-point function, we 
address the analytic continuation
to Minkowski space.

The first point to note is that 
the nucleon energy in an external Euclidean 
electric field is not directly related to that
in a magnetic field.
This should be obvious from Eq.~\eqref{eq:nukeEinB}, 
for example, there is no nucleon electric dipole moment. 
At a deeper level, while the gauge fields only differ
by a naming of the coordinates, 
the nucleon correlation functions
require us to identify single particle contributions. 
In the presence of background magnetic and electric fields, 
the pion fluctuations cohere differently in time to form a nucleon, 
and this is borne in upon LSZ reduction. 
We must thus perform a separate calculation to arrive
at the electric field result.

There are no local contributions
to the nucleon two-point function
in an electric field until $\cO(\e^3)$. 
The loop diagrams contributing to the 
electric-field dependence of the 
nucleon two-point function to $\cO(\e^2)$ 
have been depicted above in Figure~\ref{f:nucleon}.
Diagrams $A_1$ and $A_2$ vanish; 
$A_2$ for precisely the same reason as with 
the magnetic case, $v_\mu A_\mu = 0$ in our gauge. 
The derivative, $v_\mu \partial_\mu$, acting inside the loop in $A_1$
does not bring down a power 
of good momentum but the Gau\ss ian integral 
is odd after a shift of integration variable. 
Diagrams $B_2$ and $B_3$ are also simple to evaluate. 
Insertion of the gauge potential
produces the time at the vertex. 
Upon amputation of the external propagators, the
vertex times are set equal to the times of the source and sink
(and only the relative time enters the meson propagator).
Locating the source at $t = 0$ forces these diagrams to vanish. 
The remaining two diagrams, $B_1$ and $C$,
yield non-vanishing contributions to the nucleon two-point 
function.
Their evaluation and renormalization have been 
relegated to Appendix~\ref{s:nukeEcalc}.

The complete expression for the nucleon energy%
\footnote{
In evaluating the heavy nucleon two-point function in 
an electric field at this order, we can still speak of the nucleon
energy. For the proton, in strong fields we must ultimately 
sum the Born couplings to the charge. In that case, we 
can still speak of an effective energy, as was the case with 
the charged pion above. 
} 
in a Euclidean electric field $\cE$ to $\cO(\e^2)$ is given by
\begin{eqnarray} \label{eq:nukeEinE}
E_N(\cE) 
&=& 
M_N
+ g_A^2 \, \d E(0, \cE^2)
+ \frac{8}{9}  g_{\D N}^2 \, \d E(\D, \cE^2)
.\end{eqnarray}
The function $\d E(\D, \cE^2)$ represents the non-analytic contributions from 
loop diagrams, 
and has the explicit form
\begin{eqnarray}
\d E(\D,\cE^2)
&=&
\frac{- \sqrt{\pi}}{ 2 ( 4 \pi f)^2}
\int_0^\infty \frac{ds}{s^2} e^{ - s m_\pi^2}
\Bigg\{ \sqrt{\frac{2 \cE}{\sinh 2 \cE s}} e^{\frac{\D^2}{ \cE \coth \cE s}} \Erfc \left( \frac{\D}{\sqrt{\cE \coth \cE s}} \right)
[ 2 + \cE s \coth \cE s] 
\notag \\
&& \phantom{space}
- 
\frac{\cE^2 s}{\cosh \cE s \sqrt{\cE \coth \cE s}}
\left[ 
1 + \frac{2 \D^2}{ \cE \coth \cE s}
\right]
e^{\frac{\D^2}{\cE \coth \cE s}}
\Erfc \left( \frac{\D}{\sqrt{\cE \coth \cE s}}\right)
\notag \\
&& \phantom{space}
+
\frac{2 \D }{\sqrt{\pi}}
\frac{\cE s }{\cosh \cE s \coth \cE s}
- \frac{3 e^{s \D^2}}{\sqrt{s}}  \Erfc \left( \sqrt{s} \D \right)
\Bigg\} \label{eq:gawdawfulmess}
.\end{eqnarray}
Series expanding the integrand in powers of the field $\cE$, we arrive 
at the asymptotic series
\begin{equation}
\d E(\D,\cE^2) = \sum_{n=1}^\infty \d E^{(2n)}(\D) \, \cE^{2n}
.\end{equation}
From this expansion, we can read off the terms in the 
asymptotic expansion in Minkowski space
\begin{equation} \label{eq:easycont}
\d E(\D, - E^2) = \sum_{n=1}^\infty (-1)^{n} \d E^{(2n)}(\D) \, E^{2n}
.\end{equation}
Comparing with the general form of the energy in a weak electric field,
\begin{equation}
E_N = M_N - \frac{1}{2} 4 \pi \a_E \, E^2 + \ldots
,\end{equation}
we can identify the electric polarizability $\a_E$ as 
the first term in the asymptotic expansion, namely
\begin{equation}
\a_E =  \frac{1}{2 \pi} \left[  g_A^2 \d E^{(2)}(0) + \frac{8}{9} g_{\D N}^2 \d E^{(2)}(\D) \right]
.\end{equation}
Deriving the first coefficient in the asymptotic series from Eq.~\eqref{eq:gawdawfulmess}, 
we find
\begin{equation}
\d E^{(2)}(\D) = 
\frac{1}{6 (4 \pi f)^2}
\left[
\frac{9 \D}{\D^2 - m_\pi^2}
- 
\frac{\D^2 - 10 m_\pi^2}{2 ( \D^2 - m_\pi^2)^{3/2}}
\log 
\left( 
\frac{\D - \sqrt{\D^2 - m_\pi^2}}{\D + \sqrt{\D^2 - m_\pi^2}}
 \right)
\right]
\label{eq:nukeLOE}
,\end{equation}
from which it follows that $\d E^{(2)}(0) =  5  / (96  \pi f^2 m_\pi)$.
From this term in the expansion, we deduce
the known one-loop value for the nucleon electric 
polarizability~\cite{Bernard:1991rq,Bernard:1991ru,Hemmert:1996rw}. 
Notice this first coefficient in the weak field expansion 
is singular in the chiral limit. 
In general, the behavior of the coefficients near the chiral limit is:
$\d E^{(2n)}(\D) \sim m_\pi^{3 - 4 n}$.
Working to one order higher in the series, we find
the coefficient of the fourth-order term
\begin{eqnarray}
\d E^{(4)} (\D) 
&=&
\frac{1}{\pi^2 f^2  (\D^2 - m_\pi^2)^4} 
\Bigg[
\D 
\frac{ 14 \D^6 - 227 \D^4 m_\pi^2 + 2932 \D^2 m_\pi^4 + 5156 m_\pi^6}
{23040 m_\pi^4}
\notag \\
&& \phantom{spacer}
+
\frac{53 \D^4 + 412 \D^2 m_\pi^2 + 60 m_\pi^4}{3072 \sqrt{\D^2 - m_\pi^2}}
\log 
\left( 
\frac{\D - \sqrt{\D^2 - m_\pi^2}}{\D + \sqrt{\D^2 - m_\pi^2}}
 \right)
\Bigg]
\label{eq:nukeNLOE}
,\end{eqnarray}
from which follows the value $\d E^{(4)}(0) = - 5 / (256 \pi f^2 m_\pi^5)$.

%
\begin{figure}
\epsfig{file=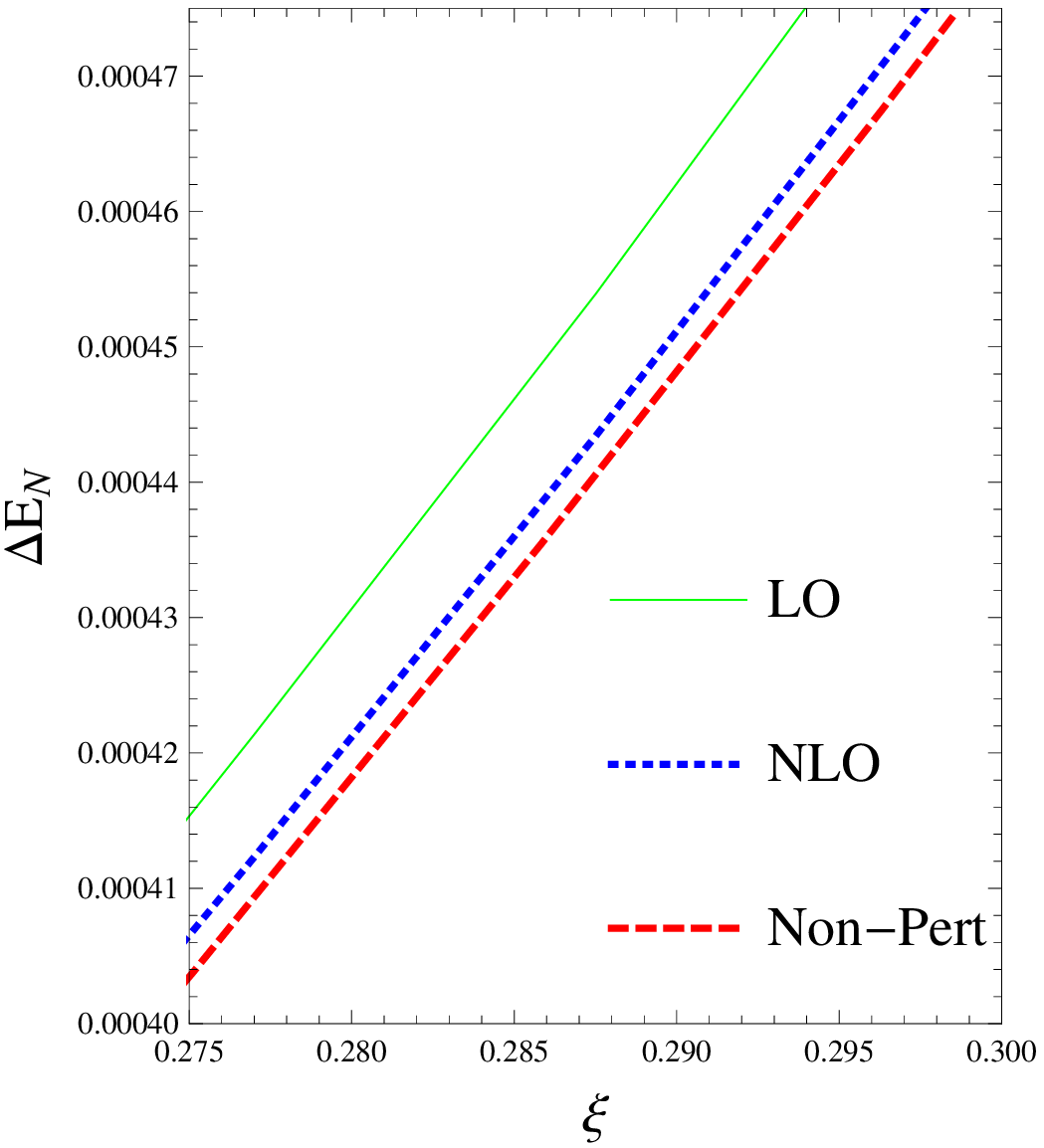,width=5.75cm} 
\epsfig{file=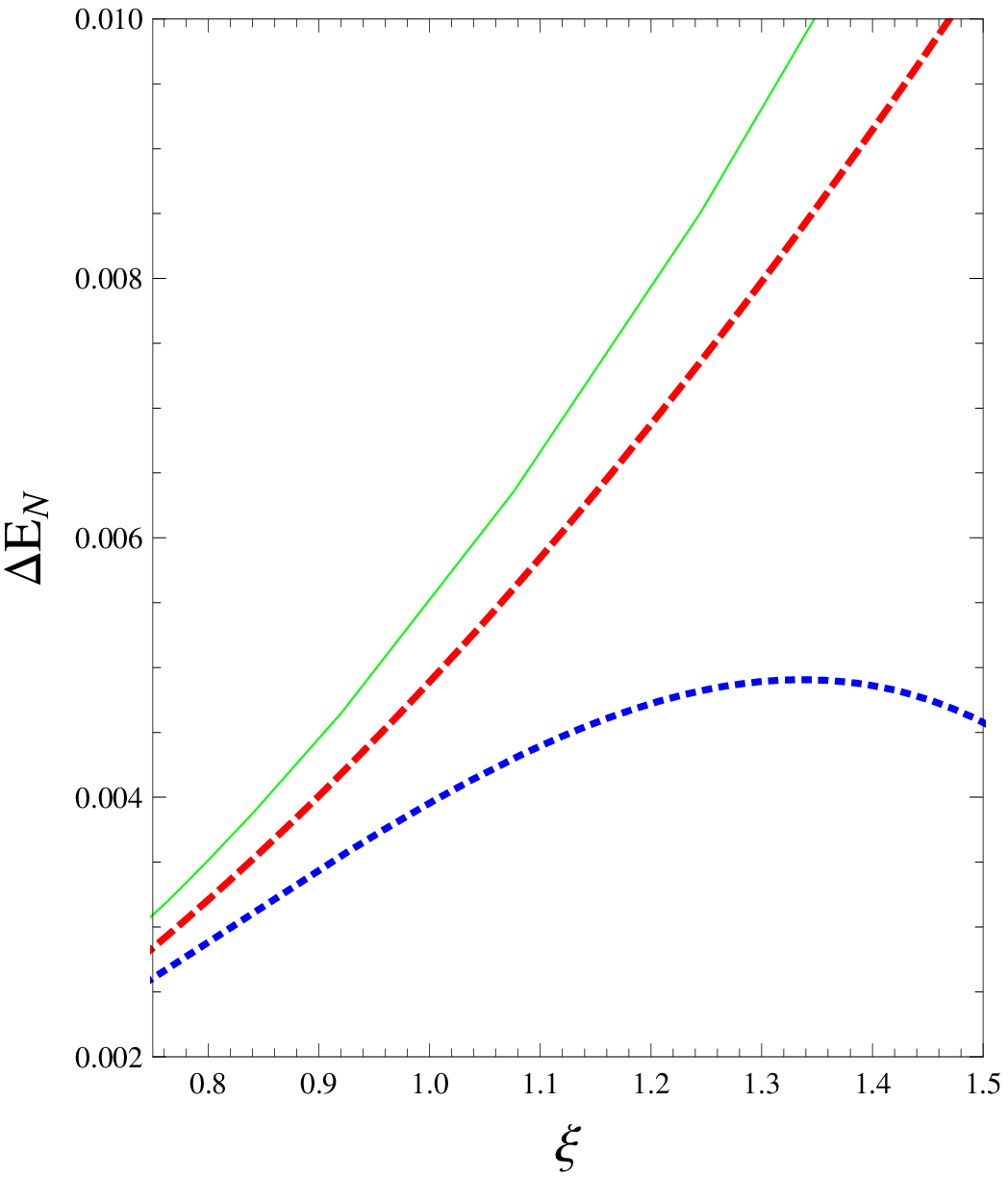,width=5.5cm} 
        \caption{Comparison of perturbative and non-perturbative results for the    
         nucleon energy in constant Euclidean electric fields. 
         Plotted is the non-perturbative result for $\D E_N \equiv \frac{E_N(\cE) - M_N}{M_N}$, 
         with $E_N(\cE)$ given in Eq.~\eqref{eq:nukeEinE}, 
        as well as the LO and NLO weak-field approximations to this result. 
        The expansion parameter $\xi = e |\cE| / m_\pi^2$.
        }
	\label{f:nukeE}
	\end{figure}
%

From these first few terms in the perturbative expansion in field strength, 
we can contrast with the non-perturbative result in Eq.~\eqref{eq:nukeEinE}. 
This comparison with the LO result from Eq.~\eqref{eq:nukeLOE}, and NLO 
result from both Eqs.~\eqref{eq:nukeLOE} and \eqref{eq:nukeNLOE} 
is shown in Figure~\ref{f:nukeE}. 
We show two distinct regions for the expansion parameter $\xi = e |\cE| / m_\pi^2$.
When $\xi \lesssim 1/3$, the perturbative expansion is under control. 
For $\xi \gtrsim 1$, the expansion has broken down, and the LO result
give the best approximation to the non-perturbative answer.

It remains to consider the analytic continuation necessary to arrive at the 
nucleon energy in an electric field $E$. 
The perturbative expansion has been trivially continued in 
Eq.~\eqref{eq:easycont}.
Unlike the pion case, however, the non-perturbative result for
$E_N(\cE)$
has a rather complicated structure in the complex plane. 
The analytic continuation can be performed using a trick,
as detailed in Appendix~\ref{s:analcont}. 
As the resulting expressions are rather cumbersome,
we restrict our attention to the case $\D = 0$. 
In that case, we have from Eq.~\eqref{eq:evenmoregawdawfulmess}
\begin{eqnarray}
\Re \mathfrak{e} 
\Big[
\d E(0, - E^2)
\Big]
&=&
\sqrt{\frac{\pi}{2}}
\frac{ |e E|^{3/2}}{ 2 ( 4 \pi f)^2}
\int_0^\infty \frac{ds}{s^2}
\left[
\cos \left( \frac{s m_\pi^2}{e |E|} \right)
+
\sin \left( \frac{s m_\pi^2}{e |E|} \right)
\right]
f(s),
\\
\Im \mathfrak{m} 
\Big[
\d E(0, - E^2)
\Big]
&=&
\sqrt{\frac{\pi}{2}}
\frac{ |e E|^{3/2}}{ 2 ( 4 \pi f)^2}
\int_0^\infty \frac{ds}{s^2}
\left[
\cos \left( \frac{s m_\pi^2}{e |E|} \right)
- 
\sin \left( \frac{s m_\pi^2}{e |E|} \right)
\right]
f(s)
,\end{eqnarray}
with
\begin{equation}
f(s) 
= 
\sqrt{\frac{2}{\sinh 2 s}}
( 2 + s \coth s) 
- 
\frac{s}{\cosh s \sqrt{\coth s}}
- 
\frac{3}{\sqrt{s}}
.\end{equation}
From the imaginary part of $\d E(\D, -E^2)$, 
one can deduce the total nucleon 
decay-width in an electric field. 
Here $p$ ($n$) decays to $\pi^+ n$ ($\pi^- p$)
plus any number of $\pi^+ \pi^-$ pairs.

%
\begin{figure}
\epsfig{file=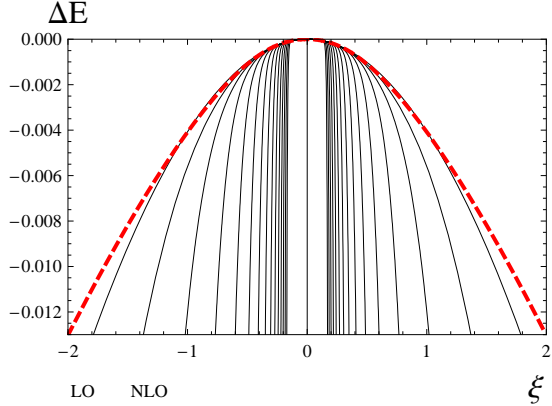,width=8cm} 
        \caption{Further comparison of perturbative and non-perturbative results for the nucleon in constant electric fields. 
        The real part of the non-perturbative energy shift (for $g_{\D N} = 0$) is plotted along with the first fifteen perturbative corrections
        in the strength of the field squared. 
        The expansion parameter $\xi = e |E| / m_\pi^2$. 
        }
	\label{f:Reel}
\end{figure}
%

In Figure~\ref{f:Reel}, we compare results in Minkowski space for the shift in nucleon energy
as a function of the expansion parameter $\xi = e |E| / m_\pi^2$. 
We plot the real part of the non-perturbative result for the relative shift in energy 
$\D E = \frac{E_N(E) - M_N}{M_N}$.
For simplicity, we have artificially set the nucleon-delta axial coupling to zero,  $g_{\D N} = 0$,  to exclude contributions from
virtual deltas. 
With this simplification, we can easily derive higher-order terms 
in the perturbative expansion, Eq.~\eqref{eq:easycont}, and have
done so up to fifteenth order, $n=15$, in the figure.
The higher-order results accumulate near $\xi = 0$, which is by now familiar behavior. 
Additionally the LO shift in nucleon energy proportional to $E^2$ does well to describe the non-perturbative
result not only qualitatively, but sometimes quantitatively, beyond the perturbative regime.

%
\section{Conclusion}
\label{s:summy}

Above we have investigated the effects of strong electric and magnetic 
fields on hadrons using chiral perturbation theory. 
We worked with a power counting that treats 
the strength of the external field non-perturbatively,
while maintaining the perturbative chiral expansion.
The field strengths meeting this criterion 
are  
$B \lesssim 10^{15} \, \texttt{T}$
for magnetic fields, and
$E \lesssim 10^{24} \, \texttt{V/m}$
for electric fields.

For fields in this regime, 
we derived the effective mass
of the charged and neutral pions.
Fortuitous cancelations at one loop
lead to only a quadratic shift of the 
charged pion effective mass. 
This shift encompasses the 
electromagnetic polarizability. 
For the neutral pion, loop graphs
give rise to an effective mass with 
non-analytic dependence on the 
external field. 
The neutral pion polarizability is the 
first field-dependent term in the 
asymptotic expansion of the non-perturbative 
result. 
The calculation of the pion effective mass
was easily carried out for a Euclidean electric field,
owing to simplifying topological features of the one-loop graphs. 
These results were then analytically continued
to Minkowski space. 
Asymptotic expansions were explored in the weak 
field limit, showing remarkable agreement between 
the non-perturbative result and the leading-order contribution
well beyond the range where the perturbative expansion in field strength
breaks down. 
This agreement, however, cannot be improved upon
by higher-order terms in the series.

We considered the effects of strong fields on the nucleon by 
using heavy baryon chiral perturbation theory, and including
the nearby delta resonances. 
In magnetic fields, we derived the shift of the nucleon 
energy accounting for both spin and isospin dependence. 
Expressions for polarized and unpolarized energy shifts 
that are non-perturbative in the field strength were derived, and 
compared to their perturbative weak-field limits.
Chiral analysis confirmed that for sufficiently strong magnetic
fields the proton becomes unstable to beta-decay. 
The energy difference between ground-state nucleons  
was demonstrated to be largely insensitive to non-perturbative 
effects from the external field. 
The first chiral corrections to the ground-state energy difference
occur at one order higher than we worked. 
It would be interesting to determine such corrections.
Finally we considered the effects of a strong electric field 
on the nucleon. 
An intricate calculation using the coordinate-space LSZ
reduction allowed us to calculate the field-dependent 
shift in nucleon energy.
The result exhibits considerably complicated analytic structure. 
We compared the non-perturbative Euclidean space result
to the weak field expansion. 
Using a novel trick, we were able to analytically continue
the nucleon result from a Euclidean electric field to 
Minkowski space. 
The real and imaginary contributions to the nucleon energy
were expressed in terms of well-defined one-dimensional 
integrals, albeit still fairly complicated. 
 Simplifying to the case without delta loop contributions, 
 we then compared the real part of the energy shift 
 to the perturbative expansion in Minkowski space. 
Similar to the pion case, the leading-order term in the 
perturbative expansion in field strength agrees well with the non-peturbative 
result when perturbation theory does not apply. 
Again improvement is not possible: higher-order terms in the strength of the field necessarily
spoil the agreement.

A practical application of our work concerns background 
field measurements
of magnetic moments and electromagnetic polarizabilities 
using lattice QCD simulations. 
In the background field approach, one simulates
QCD with classical electromagnetic fields.
The hadron spectrum is measured at various
strengths of the external field.
The measured energy shifts are then used
to deduce electromagnetic properties. 
On a torus, constant fields must satisfy 
quantization conditions which accordingly restrict
the available values of field strengths. 
On current-size lattices, the allowed field strengths
are rather large. Furthermore as pion masses are brought down
to the physical point,
the expansion parameter, $ e |B| / m_\pi^2$ or $e |\cE| / m_\pi^2$, 
will approach or even exceed unity. 
It is thus important to understand non-perturbative
effects the external field has on QCD bound states. 
Knowledge of higher-order terms in the field strength, 
and their behavior in the chiral limit,
will additionally aid in a cleaner extraction of 
magnetic moments and electromagnetic polarizabilities
from lattice data.

Finally based on our study, one can make 
some rather speculative observations
about perturbative expansions.
There are well-known cases where perturbation 
theory is applied in regimes of questionable validity. 
For example, $SU(3)$ chiral perturbation theory
for baryons%
\footnote{
The meson sector is not necessarily
perturbative either: the expansion parameter
$\e^2 \sim m^2_\eta / \L_\chi^2 \approx 1/4$.
}
has an expansion parameter 
of questionable smallness,
$\varepsilon \sim m_\eta /  M_B \approx 1/2$.
The leading $SU(3)$ corrections to the octet
baryon masses lead to the 
Gell-Mann--Okubo formula,
which experimentally holds to less than one percent. 
Could this auspicious result be the consequence, 
not of a controlled expansion about the $SU(3)$ chiral limit, 
but of a perturbative expansion that has already 
broken down?
In considering strong external electric and magnetic fields,
we have seen a few examples of leading-order 
terms in remarkable agreement with non-perturbative results.
This agreement, of course, must be observable dependent because
such behavior is not characteristic of all asymptotic expansions. 
If $SU(3)$ chiral perturbation theory has broken down 
at the physical strange quark mass, we might expect some
leading-order results to work well, while others
would show uncontrolled behavior. 
Phenomenologically it is rather difficult to assess
the convergence of the three-flavor expansion, but with 
lattice QCD data, we may soon have new insight.

Lattice QCD extrapolations themselves provide another 
example where the applicability of a perturbative expansion is questionable:
how light must the pion mass be in order to extrapolate using chiral perturbation theory 
to the physical point?
Interesting lattice results have been obtained for  
$\pi \pi$,  $\pi K$, and $K K$ 
scattering lengths~\cite{Beane:2005rj,Beane:2006gj,Beane:2007xs,Beane:2007uh}. 
The lattice-determined scattering lengths at values of the pion mass
ranging between two and three times the physical value appear to be in remarkable 
agreement with current-algebra predictions made long ago by Weinberg~\cite{Weinberg:1966kf}. 
Viewed from a modern standpoint, 
such predictions are the leading-order contributions 
from tree-level graphs in chiral perturbation theory. 
We might speculate that this tenacious agreement with leading-order 
chiral perturbation theory arises, 
not from surprisingly small next-to-leading order corrections, 
but from the pion masses being too large for the perturbative expansion
to be reliable.
Ultimately lattice QCD simulations nearing the physical point
will shed light on the regime of validity of chiral perturbation theory. 
Nonetheless, the study of hadrons in strong electric and magnetic fields
gives us an exactly soluble framework for studying non-perturbative 
effects, and gaining insight in to the behavior of perturbation theory.

\begin{acknowledgments}
This work is supported in part by the 
U.S.~Department of Energy,
under
Grant No.~DE-FG02-93ER-40762.
\end{acknowledgments}

\appendix

\section{Technical Details: Nucleon in Magnetic Fields }
\label{s:nukeBcalc}

In the main text, we argued that the only contributing diagram 
to the nucleon propagator in an external magnetic field 
is that of topology $C$ in Figure~\ref{f:nucleon}. 
Here we detail the calculation of this diagram, 
and its renormalization.

Let $G(x,0)$ denote the full nucleon two-point function, 
and $D_\cB(x,0)$ the free heavy baryon propagator.
The contribution 
$\delta G(x,0)$ 
from the sunset diagram with an internal nucleon has the form
\begin{eqnarray}
\delta G(x,0) 
=
C \int d^4 y \int d^4 z 
D_N(x,y) S \cdot \overset{\rightarrow}{\partial}_y D(y,z) S \cdot \overset{\leftarrow}{\partial}_z D_N(y,z) D_N(z,0)
,\end{eqnarray}
where $C$ is a flavor dependent coefficient. 
Recall $D(y,z)$ is the meson propagator in the background field, Eq.~\eqref{eq:propB}. 
We have used subscripts on partial derivatives to denote which variables they are
derivatives with respect to.
In the rest frame, the spin operators are purely spatial, and we thus need only calculate
spatial gradients of the meson propagator. 
The spin algebra reduces to spin diagonal and non-diagonal terms via the identity
\begin{equation} \label{eq:spinN}
S_i S_j = \frac{1}{4} \delta_{i j} + \frac{1}{2} i \epsilon_{ijk} S_k
.\end{equation}
A similar decomposition additionally holds for the diagram with an intermediate-state delta. 
The spin factors, however, are different owing to the spin-$3/2$ propagator
$P_{ij}$, which satisfies
\begin{equation} \label{eq:spinD}
P_{ij} = \frac{2}{3} \delta_{ij} - \frac{2}{3} i \epsilon_{ijk} S_k
.\end{equation}

The heavy baryon propagators simplify the loop integral because
the spatial coordinates, $\bm{y}$ and  $\bm{z}$, are forced to zero
(the location of the source). 
Focus first on the spin-diagonal term. 
The action of the partial derivatives on the meson propagator
produces
\begin{eqnarray} \label{eq:Cstep}
\bm{\nabla}_y \cdot \bm{\nabla}_z 
D(y,z) \Big|_{\bm{y} = \bm{z} = \bm{0}}
&=& 
\frac{1}{2}
\int_0^\infty ds 
\int 
\frac{d^{d-1} \tilde{\bm{k}}}
{( 2\pi)^{d-1}}
e^{ - i k_4 ( y_4 - z_4) }
e^{ - s E_\perp^2 / 2}
\sqrt{\frac{Q B}{2 \pi \sinh Q B s}}
\notag \\
&& \phantom{space} \times
e^{ - \a k_1^2 }
\left[
( \bm{\tilde{k}}^2  - k_4^2) +  Q^2 B^2 \a^2 k_1^2 + \frac{Q B}{\sinh Q B s}
\right],
\end{eqnarray}
where $\a = \frac{1}{Q B} \tanh \frac{1}{2} Q B s$, 
and $Q$ is the charge of the loop-meson. 
Eq.~\eqref{eq:Cstep} is an even function of $Q$, 
so we only need to distinguish between charged 
and neutral pion contributions. 
Evaluation of the momentum integrals,
with $Q$ set to unity for simplicity,
yields
\begin{eqnarray}
\bm{\nabla}_y \cdot \bm{\nabla}_z 
D(y,z) \Big|_{\bm{y} = \bm{z} = \bm{0}}
&=& 
\frac{ ( 4 \pi \mu^2)^\epsilon}{( 4 \pi)^2}
\int_0^\infty \frac{ds \,  s^{ \epsilon - 2}  \, B}{\sinh Bs}
e^{ - \frac{1}{2 s} [ ( y_4 - z_4)^2 + s^2 m^2 ] } 
\left[
s B \coth Bs
+ 
\frac{1}{2} - \epsilon
\right]
,\notag \\
\end{eqnarray}
in $d = 4 - 2 \epsilon$ dimensions. 
This result depends on the relative time 
$T$ between the vertices, 
given by 
$T= y_4 - z_4$. 
The intermediate-state baryon propagator also depends on $T$. 
Upon amputation of the diagram
(which is possible in coordinate space due to the simple form of the
heavy baryon propagator), 
this relative time turns into the time difference between source and sink.
To find the self-energy correction, we project the nucleon energy to zero 
by integrating over the relative time. 
The required relative-time integral has the form
\begin{equation} \label{eq:timeint1}
\int_0^\infty dT \, e^{- \frac{1}{2s} T^2 - \D T}
=
\sqrt{\frac{\pi s}{2}} e^{ \frac{1}{2} s \D^2} \Erfc \left( \sqrt{\frac{s}{2}} \D \right) 
,\end{equation}
in the case of a delta intermediate state.
$\Erfc(x)$ is the complement of the standard error function. 
The nucleon result follows from taking $\D \to 0$, for which $\Erfc(0) = 1$. 
To complete the evaluation of diagram $C$, we assemble the result of the loop integration
with the spin and flavor factors. The zero-field result gives the perturbative
correction to the nucleon mass. Renormalizing to the physical nucleon mass
amounts to a simple zero field subtraction. The result was given previously
as $\d E_2(B^2)$ above in Eq.~\eqref{eq:E2B}.

The non-diagonal spin term is similarly evaluated. 
The meson propagator evaluated at zero spatial position
is given by
\begin{eqnarray}
\bm{\nabla}_y \cdot \bm{\nabla}_z 
D(y,z) \Big|_{\bm{y} = \bm{z} = \bm{0}}
&=& 
Q B \sigma_3
\frac{ ( 4 \pi \mu^2)^\epsilon}{( 4 \pi)^2}
\int_0^\infty 
\frac{ ds \, s^{ \epsilon - 1} Q B}{\sinh Q B s}
e^{ - \frac{1}{2 s} [ ( y_4 - z_4)^2 + s^2 m^2 ] } 
.\end{eqnarray}
This contribution is odd with respect to the meson charge
and we must keep track of each pion contribution separately.
Amputation and projection onto zero energy proceeds just as with 
the spin-diagonal case; the relative time integration is identical. 
Lastly this contribution needs to be renormalized. 
The linear $B$ term produces the one-loop correction to the magnetic moment, 
which is subtracted by renormalizing to the physical magnetic moment. 
The resulting shift of the nucleon energy, we have denoted by 
$\d E_1(B^2)$ and is given in Eq.~\eqref{eq:E1B} above.

\section{Technical Details: Nucleon in Electric Fields }
\label{s:nukeEcalc}

In the main text, we found that the only contributing diagrams
to the nucleon propagator in an external electric field 
are $B_1$ and $C$ shown in Figure~\ref{f:nucleon}. 
Here we detail the calculation of these diagrams, 
and their renormalization.

First we handle diagram
$B_1$. 
The spin factors are easy to handle, there is only a
spin diagonal term.
The combination of vertices produces
$S_3 \, \bm{S} \cdot \bm{\nabla}$. 
The spatial gradient produces factors of good momentum
when acting on the meson propagator in Eq.~\eqref{eq:propE}. 
Thus when integrating over the momenta, 
only $\nabla_3$ can be non-vanishing because
the integrand is not a simple Gau\ss ian in $k_3$ momentum.
Hence we are lead
to the spin diagonal parts of 
Eqs.~\eqref{eq:spinN} and \eqref{eq:spinD}.
By virtue of the static approximation, the propagator is evaluated
at initial and final coordinate locations situated at the origin. 
The action of the derivative on the meson propagator produces
\begin{eqnarray}
\nabla_{y_3} 
D(y,z) \Big|_{\bm{y} = \bm{z} = \bm{0}}
&=&
\frac{1}{2} Q \cE ( y_4 + z_4)
\frac{( 4 \pi \mu^2)^\epsilon}{ (4 \pi )^2}
\int_0^\infty ds \, s^{ \epsilon - 1} e^{ - s m_\pi^2}
\frac{Q \cE}{\sinh Q \cE s}
e^{- \frac{1}{4} Q \cE ( y_4 - z_4)^2 \coth Q \cE s}
\notag \\ 
&\equiv&
\frac{1}{2} Q \cE ( y_4 + z_4)
\cD(\cE)
\label{eq:DE}
.\end{eqnarray}
We then multiply by the heavy baryon propagators, 
amputate the external legs, and integrate over the relative
time between source and sink.
The latter requires the integral
\begin{eqnarray} \label{eq:timeint2}
\int_0^\infty  
dT \,  
T^2 e^{ - \frac{1}{4} \cE T^2 \coth \cE s  - \D T}
&=&
\sqrt{\frac{\pi}{\cE \coth \cE s}}
\frac{d^2}{d \D^2}
\left[
e^{\frac{\D^2}{\cE \coth \cE s}}
\Erfc \left( \frac{\D}{\sqrt{\cE \coth \cE s}} \right)
\right]
.\end{eqnarray}
This contribution is finite and vanishes in the zero field limit,
hence there is no renormalization required.

To evaluate diagram $C$, we need to consider the action
of two factors of $\bm{S} \cdot \bm{\nabla}$ on the meson 
propagator. Because the $k_1$ and $k_2$ integrals are simple
Gau\ss ians, $\nabla_{1,2} \nabla_j$ acting in the loop is
proportional to a Kronecker-delta, for $j=1$, $2$, and $3$. 
Thus there are again only 
spin-diagonal contributions from Eqs.~\eqref{eq:spinN} and \eqref{eq:spinD}. 
The relevant contribution from the meson loop is 
\begin{eqnarray}
\bm{\nabla}_y \cdot \bm{\nabla}_z 
D(y,z) \Big|_{\bm{y} = \bm{z} = \bm{0}}
&=&
\frac{( 4 \pi \mu^2)^\epsilon}{ 2 (4 \pi )^2}
\int_0^\infty ds \, s^{ \epsilon - 2} e^{ - s m_\pi^2}
\frac{Q \cE}{\sinh Q \cE s}
e^{- \frac{1}{4} Q \cE ( y_4 - z_4)^2 \coth Q \cE s}
\notag \\
&& 
\times \left[ 2 ( 1 - \epsilon)  + Q \cE s \coth Q \cE s \right]
+ 
\frac{1}{4} Q^2 \cE^2 ( y_4 + z_4)^2 \cD(\cE)
,\end{eqnarray}
with $\cD(\cE)$ defined in Eq.~\eqref{eq:DE}.
To this expression, we append the propagators for the heavy baryons, 
amputate, and finally integrate over the time difference.
The relative time integration requires Eqs.~\eqref{eq:timeint1} and \eqref{eq:timeint2}. 
The resulting expression diverges only in the zero field limit. 
Subtracting off the divergence amounts to renormalizing the nucleon 
mass. 
Performing the subtraction 
and combining the results of diagrams $B_1$ and $C$ 
with the relevant flavor factors, we arrive at Eq.~\eqref{eq:nukeEinE}
given in the main text.

\section{Analytic Continuation}
\label{s:analcont}

In calculating hadronic two-point functions in external electric fields, 
we utilized a Euclidean electric field $\cE$.
To arrive at the Minkowski space result requires an analytic 
continuation in the field strength, $\cE \to i E$. 
For the pion case,  the requisite continuation rather fortuitously exists in 
closed form and is given by Eq.~\eqref{eq:continue}, see~\cite{Cohen:2007bt}.
The nucleon case appears considerably more complicated.
Both cases, however, can be analytically continued using the same method
which yields ordinary one-dimensional integrals.

Let us return to the expression for the neutral pion energy in a Euclidean 
electric field $\cE$. Setting the charge to unity $e=1$, we have
\begin{equation}
\cE \, \cI \left ( \frac{m_\pi^2}{\cE} \right) 
= 
\cE
\int_0^{\cE \times \infty} 
ds
\frac{e^{- \frac{s m_\pi^2}{\cE}  }}{s^2} 
\left( \frac{s}{\sinh s} - 1 \right)
.\end{equation}
The overall factor arises from the change to a dimensionless integration variable. 
Taking $\cE \to i E$, where $E$ is the electric field in Minkowski space, 
and defining $x = m_\pi^2 / E$, 
we have
\begin{eqnarray}
i \, \cI \left( - i x \right)
&=&
i  \int_0^{ i \infty} ds \frac{e^{ i  x s }}{s^2}
\left( \frac{s}{\sinh s} - 1 \right)
=
\int_{0 + i \epsilon}^{\infty + i \epsilon}
ds \frac{e^{- x s}}{s^2} \left( \frac{s}{\sin s} - 1 \right)
.\end{eqnarray}
In the last equality, we have added an $i \epsilon$ prescription
to render in the Minkowski space amplitude well defined~\cite{Schwinger:1951nm}. 
Because the integrand has only simple poles, we can evaluate
using the Cauchy principle value for the real part and 
the sum of residues for the imaginary part (technically multiplied by one-half). 
This method is not general enough for our purposes, and we pursue a different route.

By shifting the integration variable, 
all non-analyticities lie in the lower-half complex $s$-plane. 
We then consider the closed contour integral over the 
quarter circle in the first quadrant (positively oriented).
This integral is zero by Cauchy; moreover, the integral around
the quarter arc at infinity is zero by Jordan. 
Denoting the integrand by $f(s)$,  we have the equality 
\begin{equation}
\int_{0}^{\infty} ds \, f(s+ i \epsilon)
+
\int_{i \infty}^0 ds \, f(s)
=
0
.\end{equation}  
Consequently we find
\begin{equation} \label{eq:analcont}
i \cI (- i x)
=
- i \int_0^\infty 
ds \frac{e^{ - i x s}}{s^2}
\left( \frac{s}{\sinh s} - 1 \right) 
,\end{equation}
which gives a well-defined integral for the
analytically continued function. 
Calculated numerically, the real part of Eq.~\eqref{eq:analcont} 
agrees with the closed form expression given in Eq.~\eqref{eq:continue}.
The imaginary part, 
\begin{equation}
\Im \mathfrak{m} [ i \cI ( - i x ) ] 
=
- 
\int_{0}^{\infty} 
\frac{ds}{s^2} \cos (x s)
\left( \frac{s}{\sinh s} - 1 \right), 
\end{equation}
we can compute in closed form
with the aid of a Mittag-Leffler expansion
\begin{equation}
\frac{1}{s^2} \left( \frac{s}{\sinh s} - 1 \right) 
= 
2 \sum_{n=1}^\infty \frac{ (-1)^n}{s^2 + n^2 \pi^2}
.\end{equation}
Carrying out the proper time integration, leaves us with a sum
\begin{equation}
\Im \mathfrak{m} [ i \cI ( - i x ) ] 
=
\sum_{n=1}^\infty  (-1)^n  \frac{e^{ - n \pi x}}{n}
= 
-\log \left( 1 + e^{- \pi x} \right)
,\end{equation}
in agreement with the imaginary part of Eq.~\eqref{eq:continue}.

The same analysis can be applied to analytically continue
Eq.~\eqref{eq:gawdawfulmess} to determine the nucleon energy shift 
in an electric field. 
With $x = m_\pi^2 / e |E|$,
the Minkowski space shift reads
\begin{eqnarray}
\d E(\D,- E^2)
&=&
-\frac{ \sqrt{\pi} E^{3/2}}{ 2 ( 4 \pi f)^2}
\int_{0 + i \epsilon}^{\infty + i \epsilon} 
\frac{ds}{s^2} e^{ - x s }
\Bigg\{ \sqrt{\frac{2}{\sin 2 s}} e^{\frac{\D^2}{ E \cot s}} \Erfc \left( \frac{\D}{\sqrt{E \cot s}} \right)
[ 2 + s \cot s] 
\notag \\
&& \phantom{space}
+
\frac{s}{\cos s \sqrt{\cot s}}
\left[ 
1 + \frac{2 \D^2}{ E \cot s}
\right]
e^{\frac{\D^2}{E \cot s}}
\Erfc \left( \frac{\D}{\sqrt{E \cot s}}\right)
\notag \\
&& \phantom{space}
-
\frac{2 \D }{\sqrt{\pi E}}
\frac{s }{\cos s \cot s}
- \frac{3 e^{s \D^2 / E}}{\sqrt{s}}  \Erfc \left( \sqrt{\frac{s}{E}} \D \right)
\Bigg\} \label{eq:moregawdawfulmess}
.\end{eqnarray}
The integrand has poles and branch cuts.%
\footnote{
There are no essential singularities present in Eq.~\eqref{eq:moregawdawfulmess} due to the
limiting behavior of the error function:
\begin{equation} \notag
\Erfc(x) \overset{x \to \infty}{\longrightarrow} \frac{\sqrt{\pi}}{x} e^{- x^2}
.\end{equation}
} 
Each of the singularities, by virtue of the  $i \epsilon$ 
prescription, is located in the lower-half plane. 
In the case of branch cuts, we must use the principal branch 
of each function to situate the cuts lie in the lower-half plane. 
We can again consider the integral over the quarter circle in the first 
quadrant to arrive at an expression for the analytic continuation. 
Defining $y = x \D^2 / m_\pi^2$, we have
\begin{eqnarray}
\d E(\D,- E^2)
&=&
\frac{ \sqrt{i \pi} E^{3/2}}{ 2 ( 4 \pi f)^2}
\int_0^\infty \frac{ds}{s^2} e^{ -i  x s }
\Bigg\{ \sqrt{\frac{2}{\sinh 2 s}} e^{\frac{i y}{ \coth s}} \Erfc \left( \sqrt{\frac{i y}{\coth s}} \right)
[ 2 + s \coth s] 
\notag \\
&& \phantom{space}
-
\frac{s}{\cosh s \sqrt{\coth s}}
\left[ 
1 + \frac{2 i y}{\coth s}
\right]
e^{\frac{i y}{\coth s}}
\Erfc \left( \sqrt{\frac{i y}{\coth s}} \right)
\notag \\
&& \phantom{space}
+
2 \sqrt{\frac{y}{\pi}}
\frac{s }{\cosh s \coth s}
- \frac{3 e^{i y s}}{\sqrt{s}}  \Erfc \left( \sqrt{i y s} \right)
\Bigg\} \label{eq:evenmoregawdawfulmess}
.\end{eqnarray}
From this expression one can take the real and imaginary parts, although
the results are quite cumbersome to display. 
Additionally useful are expressions for the error function of complex argument.
For $a$ and $b$ real, we have
\begin{eqnarray}
\Re \mathfrak{e} 
\Big[ 
\Erfc ( a + i b)
\Big]
&=&
\Erfc (a)
- \frac{2}{\sqrt{\pi}}
e^{- a^2}
\sum_{n=0}^\infty
(-1)^n
\frac{b^{2 n + 2}}{(2 n + 2)!}
H_{2n + 1}(a)  ,
\\
\Im \mathfrak{m} 
\Big[ 
\Erfc ( a + i b)
\Big]
&=&
- \frac{2}{\sqrt{\pi}}
e^{- a^2}
\sum_{n=0}^\infty
(-1)^n
\frac{ b^{2 n + 1}}{(2 n + 1)!}
H_{2n}(a)
,\end{eqnarray}
where $H_n(a)$ are Hermite polynomials. 
Simpler expressions result in the case when $\D = 0$,
and these have been quoted in the main text.

%
%

\end{document}